# Polytypism in Few-Layer Gallium Selenide

Soo Yeon Lim,[a] Jae-Ung Lee,[a,b] Jung Hwa Kim,[c,d] Liangbo Liang,[e] Xiangru Kong,[e] Thi Thanh Huong Nguyen,[f] Zonghoon Lee,[c,d] Sunglae Cho[f], and Hyeonsik Cheong[*a]

[a]Department of Physics, Sogang University, Seoul 04107, Korea

[b]Department of Physics, Ajou university, Suwon 16499, Korea

[c]School of Materials Science and Engineering, Ulsan National Institute of Science and Technology (UNIST), Ulsan 44919, Korea

[d]Center for Multidimensional Carbon Materials, Institute for Basic Science (IBS), Ulsan 44919, Korea

[e]Center for Nanophase Materials Sciences, Oak Ridge National Laboratory, Oak Ridge, Tennessee 37831, United States

[f]Department of Physics and Energy Harvest Storage Research Center, University of Ulsan, Ulsan 44610, Korea

[*]E-mail: hcheong@sogang.ac.kr

**Abstract**

Gallium selenide (GaSe) is one of layered group-III metal monochalcogenides, which has an indirect bandgap in monolayer and direct bandgap in bulk unlike other conventional transition metal dichalcogenides (TMDs) such as $MoX_2$ and $WX_2$ (X=S and Se). Four polytypes of bulk GaSe, designated as β-, ε-, γ-, and δ-GaSe, have been reported. Since different polytypes result in different optical and electrical properties even for the same thickness, identifying the polytype is essential in utilizing this material for various optoelectronic applications. We performed polarized Raman measurement on GaSe and found different ultra-low-frequency Raman spectra of inter-layer vibrational modes even for the same thickness due to different stacking sequences of the polytypes. By comparing the ultra-low-frequency Raman spectra with theoretical calculations and high-resolution electron microscopy measurements, we established the correlation between the ultra-low-frequency Raman spectra and the stacking sequences for trilayer GaSe. We further found that the AB-type stacking is more stable than the AA′-type stacking in GaSe.

**1. Introduction**

Since graphene was first isolated in 2004,[1] two-dimensional (2D) layered materials have been studied intensely owing to the possible application of these materials in future electronics such as flexible devices. Since pristine graphene poses difficulties in using it in optoelectronic devices due to a lack of a bandgap, semiconducting transition metal dichalcogenides (TMDs) such as $MoX_2$ and $WX_2$ (X=S and Se) attract much interest and have been studied widely as alternative materials.[2–6] On the other hand, group-III metal monochalcogenides such as GaS, GaSe, and InSe are recently attracting interest as a new family of 2D layered semiconductors since they have high photo-responsivity and external quantum efficiency (EQE) in the UV-range.[7–9] GaSe is one of group-III metal monochalcogenides, with a direct bandgap energy of ~2 eV in bulk.[10,11] Additionally, GaSe has been widely used in nonlinear optical applications.[12,13] Group-III metal monochalcogenides have distinct band structure from conventional TMDs. They have a Mexican hat-shaped valence band structure around the Γ-point in momentum space, leading to an indirect bandgap in monolayer.[7,8,14–18] For example, the conduction band minimum of GaSe is at the Γ-point, but the valence band maxima are located slightly away from the Γ-point except for the bulk. However, because of the small energy difference between the valence band maxima and the Γ-point, direct transitions at the Γ-point plays an important role in the optical properties, resulting in improved optical absorption and emission.[19] Additionally, GaSe is a p-type semiconductor, which can be combined with conventional n-type TMDs.[20]

The physical properties of layered materials are sensitive to the thickness and the stacking types in addition to the properties of individual layers. Polytypism is a particular type of polymorphism in layered materials:[21,22] Even if the structures of the constituent layers are identical, different stacking between the layers in terms of relative orientations and atomic alignments result in different polytypes.[23] Because many physical properties depend on the polytype, it is important to differentiate different polytypes in layered materials. Especially, optoelectronic features such as band-gap tunability or valley polarization can be manipulated by controlling the stacking sequence. For example, ABA- and ABC-stacked trilayer graphene exhibit very different band structures.[24–26] Raman spectroscopy is widely used as a tool to identify the stacking sequences because it is fast, reliable, and easy for identifying polytypes compared to other tools.[27–31] For example, there exist two types of stacking sequences in $MoS_2$, 3R and 2H, which can be easily identified by using certain low-frequency Raman modes.[32,33]

Monolayer group-III metal monochalcogenides consists of covalently bonded two post transition metal atoms (M) and two chalcogen atoms (X) (X-M-M-X): the monolayer GaSe consists of two Ga atoms sandwiched between two Se atoms as shown in Fig. 1a, whereas it has a hexagonal structure in the top view like conventional TMDs. An individual layer of GaSe has the $D_{3h}$ symmetry with four atoms in the unit cell. Bulk GaSe has four different polytypes, designated as $\beta$ (2H)-, $\varepsilon$ (2H′)-, $\gamma$ (3R)-, and $\delta$ (2H-3R)-GaSe as shown in Fig. 1b. Each polytype corresponds to AA′AA′AA′…, ABABAB…, ABCABC… and AA′B′BAA′B′B… stacking sequences, respectively.[34] The position B indicates translation of the top layer A by one-third of a unit cell along the armchair direction of the hexagonal lattice. The position C is an equivalent translation of the position B. The position A′ and B′ denote the mirror images of the positions A and B, respectively, with respect to the plane bisecting an armchair bond. All the Ga atoms in one layer are over Se atoms in the successive layer for the AA′ stacking without centered atoms in hexagons, whereas Ga or Se atoms are over the hexagon centers for the AB stacking. The $\delta$-GaSe is a mixed type of AA′ and AB stackings (2H and 3R). The $\beta$-, $\varepsilon$-, $\gamma$-, and $\delta$-GaSe have space groups of $D_{6h}^4$, $D_{3h}^1$, $C_{3v}^5$, and $C_{6v}^4$, respectively.[33] The $\beta$-, $\varepsilon$-, and $\delta$-GaSe have hexagonal structures whereas the $\gamma$-GaSe has a rhombohedral structure. In the bulk phase, the $\varepsilon$-GaSe polytype has been most extensively studied, followed by the $\gamma$-GaSe polytype.[35–44] Since the unit cell of the $\delta$-GaSe contains four layers, the Brillouin zone of the $\delta$-polytype is smaller than the others.[43]

In this work, we carried out polarized Raman spectroscopy of exfoliated few-layer GaSe samples and identified several different Raman spectra for the same thickness. In particular, we identified 4 types of ultra-low-frequency Raman spectra for trilayer GaSe. By comparing with high resolution (scanning) transmission electron microscopy (HR-S/TEM) results and theoretical calculations, we establish the correspondence between the Raman spectra and the specific stacking sequences.

2. Methods

2.1. Synthesis of GaSe crystal

GaSe single crystals were successfully grown by temperature gradient method (TGM). Firstly, high purity (99.999%) Ga and Se powders were prepared in a stoichiometric ratio for the growth process, then loaded into quartz tubes which have a cylindrical shape with a conical bottom. The tubes along with powders were evacuated to an atmosphere of $10^{-4}$ Torr and sealed by oxygen-hydrogen flame. This tube was contained and subsequently sealed under vacuum by another quartz tube in order to protect it from ambient in case the inner tube breaks due to the high vapor pressure of Se during heating process or due to the difference in thermal expansion coefficient of GaSe and quartz tube during cooling process. The ampules were placed into a vertical furnace and gradually heated to 975 °C, about 15 °C above the melting point of GaSe, then maintained at that temperature for 16h for making compound. After that, the molten material was cooled down below the melting point at a very low rate of about 1 °C per hour. The growth process required two weeks to finish and acquire samples.

2.2 Sample transfer and exfoliation

The samples were fabricated on $SiO_2$/Si substrates with 90 nm or 285 nm oxide layer by mechanical exfoliation from GaSe bulk crystal. For TEM measurements, the samples were exfoliated on the polydimethylsiloxane (PDMS) and transferred to graphene on $SiO_2$/Si substrate.

We kept the samples in vacuum to avoid degradation from air exposure (see Fig. S1†). The thickness of the sample was confirmed by AFM (NT-MDT).

**2.3 Raman measurements**

We performed Raman spectroscopy measurements mainly with a diode-pumped solid-state laser with the wavelength of 532 nm (2.33 eV) and the power of ~0.1 mW. In a separate set of measurements, we found that the sample degradation is kept minimal at this excitation power (see Fig. S1†). A 40× objective lens (N.A.=0.6) was used to focus the laser to a spot of a ~1 μm diameter and also collected the scattered light from the sample. The scattered light from the sample was dispersed with a Jobin-Yvon Horiba iHR550 spectrometer (2400 grooves/mm) and was detected with a charge-coupled-device (CCD) using liquid nitrogen for cooling. Reflective volume Bragg gratings (OptiGrate) were used as notch filters to remove Rayleigh scattered signal, which enables us to observe Stokes and anti-Stokes Raman bands down to 5 cm$^{-1}$. Polarized Raman measurements were performed with polarizers, λ/4 waveplate, and λ/2 wave plates for selecting appropriate polarizations of incident and scattered light.

**2.4 TEM measurements**

For matching between Raman and TEM results, S/TEM analysis was performed with the particular flakes, characterized by Raman spectroscopy. For S/TEM analysis, exfoliated GaSe flakes on the SiO$_2$/Si substrate were transferred onto a TEM grid by a direct transfer. Because GaSe is significantly oxidized in ambient condition, transfer should be completed in short time. Direct transfer has advantages on reduced transfer time and clean without poly(methyl methacrylate) (PMMA) residue. Exfoliated GaSe flakes were analyzed using an aberration-corrected FEI Titan cube G2 60-300 with monochromator. Atomic resolution S/TEM was applied for the analysis of definite stacking order of trilayer GaSe. TEM image simulation was implemented in MacTempasX for interpreting exact stacking order. All S/TEM analysis was operated at 80 kV.

**2.5 Theoretical calculations**

Plane-wave density functional theory (DFT) calculations were carried out using the VASP package,[45] with the projector augmented-wave (PAW) method utilized for electron-ion interactions and local density approximation (LDA) for exchange-correlation interactions. For bulk GaSe, both atomic positions and cell volumes were allowed to relax until the residual forces were below 0.001 eV Å$^{-1}$, where we adopted a k-point sampling of 18×18×4 in the Gamma-centered Monkhorst-Pack scheme[46] with the energy cutoff set at 350 eV. Then, bilayer and trilayer GaSe systems at various stacking configurations were modeled by a periodic slab geometry based on the optimized bulk structure. A vacuum separation of 22 Å in the out-of-plane direction was used to avoid spurious interactions with the periodic replicas. For the 2D slab calculations, all atoms were relaxed until the residual forces were also below 0.001 eV Å$^{-1}$, with the k-point sampling of 18×18×1 and the energy cutoff of 350 eV. Subsequently, Raman spectra (both phonon frequencies and Raman intensities) were calculated based on the fully relaxed geometries, by computing the dynamic matrix and derivatives of the dielectric tensors with respect to phonon vibrations.[47,48] Specifically, the dynamic matrix was calculated using the *ab initio* direct method implemented in the PHONON software.[49] In the finite difference scheme, the Hellmann-Feynman forces in the 3×3×1 supercell were computed by VASP for both positive and negative atomic displacements (δ = 0.03 Å), and used in PHONON to construct the dynamic matrix, whose diagonalization provides phonon frequencies and phonon eigenvectors (i.e., vibrations). The derivatives of the dielectric tensor were also calculated by the finite difference approach. For both positive and negative atomic displacements in the single unit cell ($\delta = 0.03$ Å), the dielectric tensors were computed by VASP using density functional perturbation theory and then imported into PHONON to generate their derivatives.[47,48] Finally, Raman intensity of every phonon mode was obtained for a given laser polarization set-up in the typical experimental back scattering configuration to yield Raman spectra after Lorentzian broadening.

## 3. Results and discussion

The few-layer GaSe samples were prepared by mechanically exfoliating from bulk crystal flakes grown by the temperature gradient method (see Methods section for details) onto $SiO_2$/Si substrates with a 90 or 280 nm-thick oxide layer. Fig. 1c and d show the optical image and the corresponding atomic force microscope (AFM) image of a sample, respectively. The line scan confirmed the thickness of the monolayer sample as shown in Fig. 1e. The monolayer, which is barely resolved in the optical image but can be identified in the AFM image, has a thickness of ~1 nm which is consistent with the inter-layer periodicity of ~0.8 nm from x-ray diffraction measurements.[50,51]

Fig. 2a shows polarized Raman spectra of monolayer (1L) to five-layer (5L) and bulk GaSe, measured with the 532-nm laser as the excitation source. The stacking sequences of the samples are not identified here. The polarized Raman spectra were obtained in parallel [$\bar{z}(xx)z$] and cross [$\bar{z}(xy)z$] polarization configurations. In the parallel polarization configuration, the polarization directions of the incident light and the scattered light are parallel to each other. In the cross configuration, the polarization directions of the incident light and the scattered light are perpendicular to each other. The peaks at ~59, 134, 213, and 308 cm$^{-1}$ correspond the intra-layer $E''$, $A'_1(1)$, $E'$, and $A'_1(2)$ modes of GaSe, respectively, which are similar to recently reported results.[52] It should be noted that the notations of the vibrational modes depend on the number of layers, but we will use the corresponding notations for bulk ε-GaSe unless noted otherwise. The Raman intensities of the $A'_1(1)$, $E'$, and $A'_1(2)$ modes measured in the parallel polarization configuration are not dependent on the incident polarization direction due to the isotropic in-plane symmetry of GaSe (see Fig. S3†). The peak at ~59 cm$^{-1}$ is very weak and can be clearly seen only for thick samples. In addition, weak peaks are occasionally observed at ~245 and 251 cm$^{-1}$ in comparatively thick samples (see Fig. S4†). Their origin is not entirely clear and there are several explanations for them. They could be the E′(LO) mode from ε-GaSe, the $E^2_{1g}$ mode from β-GaSe, or the E mode from γ-GaSe.[37–39,41,42,53] On the other hand, our theoretical calculation predicts that a phonon mode corresponding to the bulk Raman inactive $A''_2$ mode is located near 250 cm$^{-1}$, and can be Raman activated and appear in few-layer samples due to the reduction in symmetry. This is a common phenomenon in other TMDs such as MoTe$_2$.[54] Also, some forbidden modes may appear due to symmetry breaking from crystal imperfections, and even amorphous Se with the Raman peak at ~250 cm$^{-1}$ may exist on the surface since GaSe is easily oxidized.[55–58] Since these peaks are very weak and clearly observed only in relatively thick samples, we will not discuss them further in this work. In the ultra-low-frequency range below 30 cm$^{-1}$, there are several modes due to the inter-layer shear and breathing vibrations. These are acoustic-like vibrations of the entire layer against each other in the direction parallel to the layer plane (shear) or perpendicular to the layer (breathing).[43,59,60] Because the number of these inter-layer modes and their frequencies depend sensitively on the number of layers, they are the most reliable fingerprints of the number of layers. Furthermore, the inter-layer modes are known to be sensitively depend on stacking sequences in other layered 2D materials such as $MoS_2$.[31–33,61,62]

Group theory predicts that the in-plane vibrating E modes (including the shear modes) are observed in both the parallel and cross polarization configurations whereas the out-of-plane vibrating A modes (including the breathing modes) are observed only in the parallel polarization configuration (see Note S1†). Since the breathing modes often overlap with the shear modes, it is difficult to resolve the breathing modes in the linearly polarized Raman measurements. However, if one uses circularly polarized light, the breathing modes are allowed only in the same circular polarizations of the incident and scattered photons, whereas the shear modes are allowed only when the circular polarizations of the incident and scattered photons are opposite (see Note S1†).

Fig. 2b shows the peak positions of the three stronger intra-layer modes as a function of the number of layers. Whereas the higher frequency $E'$, and $A'_1(2)$ peaks show little variation with the number of layers, the $A'_1(1)$ peak shows a monotonic blueshift with the number of layers which can be used to determine the thickness for thin samples. Fig. 2c shows the positions of the inter-layer vibration modes. The positions of the shear modes were determined from the (linearly) cross-polarization configuration spectra. The breathing modes were determined from the circularly polarized measurements, but their weak intensities prevented us from determining the positions reliably except for a few cases. We, therefore, focus on the shear modes in our analyses below. For each thickness, we measured several samples and found that the shear mode positions vary within experimental uncertainty. Furthermore, although the relative intensities of the shear modes vary for different stacking sequences, the peak positions do not show measurable differences. However, the peak positions of the shear modes show strong variations with the sample thickness in Fig. 2c, a typical behavior for inter-layer vibration modes in 2D layered materials.[63] The dashed curves are fits to the simple linear chain model in which only the layer-to-layer interaction is used as a fitting parameter. The inter-layer force constant thus determined is ~(1.33±0.03)×$10^{19}$ N m$^{-3}$ along the in-plane direction, which is smaller than those of other typical group VI and VII based TMDs layered materials.[60,64–66] Our obtained value is similar to theoretically estimated values of ~ 1.20×$10^{19}$ N m$^{-3}$ for $\beta$-GaSe and ~1.35×$10^{19}$ N m$^{-3}$ for $\varepsilon$-GaSe,[67] and our own theoretical estimation of ~ 1.33×$10^{19}$ N m$^{-3}$ for $\beta$-GaSe. Therefore, the frequencies of the shear modes shown in Fig. 2c not only allow quick and effective determination of the number of layers in GaSe samples, but also reveal that the inter-layer coupling of GaSe is weaker than many common TMDs.

Close inspection of the ultra-low-frequency Raman spectra reveals that the relative intensities of the inter-layer vibration modes vary greatly between samples of the same thickness determined by the positions of the inter-layer vibration modes and the AFM measurements. This indicates that the layers are stacked in different sequences. In order to find the correlation between the Raman spectrum and the stacking sequence, we focused on trilayer GaSe samples. For thicker samples, the number of stacking sequence variations becomes too large for a conclusive analysis. Polarized Raman measurements were performed on multiple points in different flakes (see Fig. S5†). Fig. 3a shows four typical ultra-low-frequency Raman spectra of trilayer samples measured in the cross-polarization configuration. Two shear modes at 9.6 cm$^{-1}$ ($S_1$) and 16.6 cm$^{-1}$ ($S_2$) are observed. We assigned four types of ultra-low-frequency spectra in Fig. 3a as: Type 1 (black), Type 2 (red), Type 3 (green) and Type 4 (blue). In Type 1, the intensity of the peak $S_1$ is higher than that of the peak $S_2$ whereas the intensity ratio is opposite in Type 3. In Type 2, only the peak $S_1$ appears, and in Type 4 only the peak $S_2$. The positions of the two peaks are identical in all four types as mentioned earlier. Fig. 3b shows that the $A_1'(1)$ mode frequency is also identical in all four types, indicating that the intra-layer vibration modes are not affected by the stacking sequences significantly. We examined the positions of the breathing and shear modes separately by using circularly polarized Raman measurements. We observed that the breathing mode does not show much differences in the intensity or the frequency even when the samples have different ultra-low-frequency (shear mode) spectra (see Fig. S6†), which is similar to what has been observed in twisted multilayer graphene.[68,69]

Furthermore, we investigated several trilayer flakes which show a distribution of different ultra-low-frequency spectra. Fig. 3c is an optical microscope image of one such sample, with the measurement point marked. The colors of the points match the spectrum type in Fig. 3a. Fig. 3d and e are Raman intensity images of the peaks $S_1$ and $S_2$ taken from the area indicated by the dashed box in Fig. 3c, respectively. The distribution of the peak intensities matches the spectrum types indicated in Fig. 3c, clearly showing a distribution of areas with different ultra-low-frequency Raman spectra. This result implies that GaSe exists in different areas of stacking sequences even in the same thickness flake. We also found some flakes that comprise mostly only one type (see Fig. S5e and l†). We found more flakes having various stacking sequences than flakes having only one stacking sequence, although Type 2 and Type 4 were somewhat more frequently found.

In order to correlate the Raman spectra and the stacking sequences, we theoretically calculated ultra-low-frequency Raman spectra for possible stacking sequences of trilayer GaSe. Fig. 4a shows the four different ultra-low-frequency Raman spectra calculated via first-principles density functional theory (DFT) according to the procedure explained in Method Section. The calculated non-resonant Raman intensities of low-frequency $S_1$ and $S_2$ peaks vary considerably with stacking sequences, similar to experimental trends in Fig. 3a. In contrast, Fig. 4b shows that the high-frequency intra-layer modes are virtually identical in all stacking sequences, which is reasonable because the intra-layer modes are less sensitive to stacking changes than the inter-layer modes.

In addition to the DFT method above, Raman intensities of low-frequency inter-layer modes in 2D materials can also be computed by a simple inter-layer bond polarizability model proposed in our prior work.[70] This model can provide more physical insights compared to the DFT approach. Generally speaking, Raman intensity of each normal mode is proportional to the change of the system's polarizability with respect to the normal coordinates of the corresponding vibration, and so obtaining the polarizability change by the vibration is crucial for calculating the intensity. For an inter-layer vibration mode, each layer oscillates as a quasi-rigid body, and therefore it can be treated as a single object. For the layer $i$, if the derivative of the system's polarizability with respect to its displacement is $\alpha'_i$ and its displacement during the inter-layer vibration is $\Delta r_i$, the change of the polarizability by this displacement is $\Delta \alpha_i = \alpha'_i \cdot \Delta r_i$. The total change of the system's polarizability by the inter-layer vibration is the sum of the changes of every layer: $\Delta \alpha = \sum_i \Delta \alpha_i = \sum_i \alpha'_i \cdot \Delta r_i$, where $\alpha'_i$ is related to the properties of the inter-layer bonds, including the inter-layer bond polarizabilities and the inter-layer bond vectors (lengths and directions).[70] The general form of $\alpha'_i$ can be simply determined based on the directions of the inter-layer bond vectors.[63,70] Meanwhile, the displacement of each layer, $\Delta r_i$, can be determined by the linear chain model.[65] Finally, Raman intensity of the inter-layer vibration mode is obtained based on the formula $I \propto \frac{n+1}{\omega} |\Delta \alpha|^2$, where $n = (e^{\frac{\hbar \omega}{k_B T}} - 1)^{-1}$ is the phonon occupation according to Bose–Einstein statistics and $\omega$ is the frequency of the vibration mode.

In trilayer GaSe, for an inter-layer shear vibration along the *x* direction, the polarizability change is $\Delta\alpha = \sum_i \alpha'_i \cdot \Delta x_i$, where $\alpha'_i$ can change notably with the stacking, since it is sensitive to the inter-layer bond polarizabilities and bond directions that vary with the stacking, according to the inter-layer bond polarizability model.[63,70] Similar to bilayer MoS$_2$,[63] bilayer GaSe has two stacking patterns of AA′ and AB, and the inter-layer bond properties are different as the relative layer-to-layer atomic alignments are different between AA′ and AB stackings. This is confirmed by the reported different low frequency Raman intensities of bilayer MoSe$_2$ and MoS$_2$ in these two stacking sequences.[32,33,71] For trilayer GaSe, there are a variety of stacking sequences, including AA′A, ABA, ABC, AA′B′, and A′B′B, as shown in Fig. 1b and Fig. S7.† Note that both AA′B′ and A′B′B originate from the bulk stacking AA′B′B in Fig. 1b, and A′B′B is equivalent to ABB′, where BB′ is equivalent to AA′. For AA′A stacking in trilayer GaSe, the top layer and bottom layer (i.e., layer 1 and layer 3) are in the equivalent positions, thereby giving $\alpha'_1 = \alpha'_3 = \beta_1$ and subsequently $\alpha'_2 = -2\beta_1$ (see the general relation $\alpha'_1 + \alpha'_2 + \alpha'_3 = 0$ valid for any stacking in Note S2†); for ABA stacking, the top layer and bottom layer are also in the equivalent positions, and it has the same form of inter-layer bond vectors as AA′A stacking but different inter-layer bond polarizabilities, therefore giving $\alpha'_1 = \alpha'_3 = \beta_2$ and subsequently $\alpha'_2 = -2\beta_2$ (note that $\beta_1$ and $\beta_2$ are related to the inter-layer bond polarizabilities of AA′ and AB stackings, respectively); for ABC stacking, the layer-layer stacking assumes the same AB type as ABA stacking (i.e., BC stacking equivalent to BA), but layer 2 and layer 3 have different stacking directions and thus the opposite inter-layer bond directions compared to ABA stacking, thus yielding $\alpha'_1 = -\alpha'_3 = \beta_2$ and subsequently $\alpha'_2 = 0$; for AA′B′ and A′B′B stackings, the situation is more complicated due to a mixture of AA′ stacking AB stacking, and we can derive that $(\alpha'_1, \alpha'_2, \alpha'_3) = (\beta_1, -\beta_1 + \beta_2, -\beta_2)$ for AA′B′ stacking and $(\beta_2, -\beta_2 - \beta_1, \beta_1)$ for A′B′B stacking (more details in Note S2† and our prior theory work).[70] Although the polarizability derivatives show strong dependence on the stacking sequences, the frequencies and eigenvectors (i.e., layer displacements) of inter-layer vibration modes are insensitive to the stacking patterns as demonstrated in the aforementioned experimental data. There are two inter-layer shear modes ($S_1$ and $S_2$) for the trilayer, and the normalized displacements of layer 1, layer 2 and layer 3 are $(\Delta x_1, \Delta x_2, \Delta x_3) = \frac{1}{\sqrt{2}}(1, 0, -1)$ for the lower-frequency $S_1$, and $\frac{1}{\sqrt{1.5}}(0.5, -1, 0.5)$ for the higher-frequency $S_2$, according to the linear chain model. Based on the formula $\Delta\alpha = \sum_i \alpha_i' \cdot \Delta x_i$ and $I \propto \frac{n+1}{\omega}|\Delta\alpha|^2$, we can obtain Raman intensities of the shear modes $S_1$ and $S_2$ for different stacking configurations in trilayer GaSe (detailed derivations in Note S2†):

$$\begin{aligned}
&I(AA'A, S_1) = 0; &&I(AA'A, S_2) \propto 4.68|\beta_1|^2; \\
&I(ABA, S_1) = 0; &&I(ABA, S_2) \propto 4.68|\beta_2|^2; \\
&I(ABC, S_1) \propto 4.60|\beta_2|^2; &&I(ABC, S_2) = 0; \\
&I(AA'B', S_1) \propto 1.15|\beta_1 + \beta_2|^2; &&I(AA'B', S_2) \propto 1.17|\beta_1 - \beta_2|^2; \\
&I(A'B'B, S_1) \propto 1.15|\beta_1 - \beta_2|^2; &&I(A'B'B, S_2) \propto 1.17|\beta_1 + \beta_2|^2.
\end{aligned} \quad (1)$$

From eqn. (1) it is evident that for both ABA and AA'A stacking sequences, only the higher-frequency $S_2$ mode shows non-zero Raman intensity, corresponding to the experimental Raman spectrum of Type 4 in Fig. 3a and also consistent with the DFT data in Fig. 4a. For ABC stacking, on the other hand, only the lower-frequency $S_1$ mode can be observed, consistent with the experimental Raman spectrum of Type 2 in Fig. 3a and the DFT counterpart in Fig. 4a. Such opposite trends between ABA and ABC stacking sequences are directly related to the different polarizability derivatives of layer 2 and layer 3 stemming from the opposite stacking directions and inter-layer bond directions between layer 2 and layer 3, as discussed before. It is interesting to point out that $I(ABA, S_2) \approx I(ABC, S_1)$ according to eqn. (1), in agreement with the experiment data of Type 4 and Type 2 in Fig. 3a and the DFT results in Fig. 4a.

For both AA′B′ and A′B′B stacking sequences, due to the mixture of two stacking types (AA′ and AB), it is expected that both $S_1$ and $S_2$ peaks can be detected. As the coefficients 1.15 and 1.17 in eqn. (1) are about the same, the intensity ratio between $S_1$ and $S_2$ modes is roughly $r \approx \frac{|\beta_1 + \beta_2|^2}{|\beta_1 - \beta_2|^2}$ for AA′B′ stacking, while $r' \approx \frac{|\beta_1 - \beta_2|^2}{|\beta_1 + \beta_2|^2} \approx \frac{1}{r}$ for A′B′B stacking. We note that the system's polarizability (or dielectric function) is complex, and has both real and imaginary parts due to the light absorption in experimental Raman scattering. Therefore, $\beta_1$ and $\beta_2$, parameters related to the inter-layer bond polarizabilities, are complex as well. We can define $\beta_1 = |\beta_1| e^{i\phi_1}$ and $\beta_2 = |\beta_2| e^{i\phi_2}$, where $\phi_1$ and $\phi_2$ are their phase angles, respectively. Consequently, $r = \frac{|\beta_1|^2 + |\beta_2|^2 + 2|\beta_1||\beta_2|\cos(\phi_1 - \phi_2)}{|\beta_1|^2 + |\beta_2|^2 - 2|\beta_1||\beta_2|\cos(\phi_1 - \phi_2)}$, where the difference of the phase angle $\phi_1 - \phi_2$ affects the magnitude of $r$. For non-resonant Raman scattering where the incident laser photons do not excite electrons, $\phi_1$ and $\phi_2$ should be close to zero and $\phi_1 - \phi_2$ is typically between 0° and 90°. Consequently, we have $r > 1$, and thus $I(S_1) > I(S_2)$ for AA′B′ stacking while $I(S_1) < I(S_2)$ for A′B′B stacking. This is consistent with the DFT calculations of non-resonant Raman scattering in Fig. 4a, where $I(S_1) < I(S_2)$ for A′B′B stacking. However, when the laser wavelength is near the energy of an electronic transition, $\phi_1$ and $\phi_2$ could change dramatically, leading to different intensity ratios between $S_1$ and $S_2$ modes. Therefore, there are some ambiguities in determining AA′B′ and A′B′B stacking sequences based on the Raman peak intensities of $S_1$ and $S_2$ alone.

The peak positions and the corresponding intensity ratio of peak $S_1$ and $S_2$ from our calculations (Fig. 4a and eqn. (1)) are compared with the experimental results in Fig. 3a to match the ultra-low-frequency Raman spectra to specific stacking sequences: Type 1 to AA′B′ or A′B′B, Type 2 to ABC, Type 3 to A′B′B or AA′B′, and Type 4 to ABA or AA′A. Only the Type 2 Raman spectrum can be unambiguously identified as the ABC stacking sequence, whereas there are ambiguities for the other types.

In order to identify the stacking sequences, we performed HR-S/TEM analysis on trilayer samples. For these purposes the GaSe samples were exfoliated on PDMS and transferred to mono- or bi-layer graphene on $SiO_2$/Si substrate to protect samples from etchant during the transfer process to TEM grids. After the Raman spectrum were taken, the samples were transferred onto TEM grids for HR-S/TEM analysis. We analyzed several areas of each sample in order to ascertain the repeatability. The stacking sequences are assigned by comparing the HR-S/TEM data with simulations shown in Fig. 5a to d. The simulated intensities along the red lines are shown below the simulated images. In the ABC stacking sequence, all the spots have the same intensity because each spot corresponds to 2 Ga atoms and 2 Se atoms, whereas the AA′B′, A′B′B and ABA stacking sequences have three different intensities for the spots (see Fig. S7 and table S2†). On the other hand, the AA′A stacking sequence has an in-plane hexagonal structure without centered atoms, which makes it distinct from the other stacking sequences. First of all, the Type 2 sample, which was identified as the ABC stacking sequence by comparing the Raman spectrum with the theoretical calculation, indeed shows an ABC-type HR-STEM image. The Type 4 sample, which was classified as either the ABA or AA′A stacking sequences from the Raman spectrum and Raman calculation, cannot be the AA′A stacking sequence because there is a center spot in the hexagon. Therefore, we conclude that the Type 4 corresponds to the ABA stacking sequence. The Type 3 sample can be either AA′B′ or A′B′B stacking sequences from the Raman analysis. These two types have similar HR-S/TEM images because the atomic numbers of Ga and Se are not very different: the AA′B′ stacking sequence has a spot with 2 Se atoms whereas A′B′B has a spot with 2 Ga atoms (see Fig. S7 and table S2†). From the HR-S/TEM image, the intensities of the spots have the ratio of 1:1.3:2.7, whereas the expected ratios for AA′B′ or A′B′B stacking sequences are 1:1.51:1.98 and 1:1.65:2.18, respectively (see table S2†). The experimental intensity ratio is slightly closer to the expected ratio for the A′B′B stacking sequence. Based on the analyses of the relative intensities of $S_1$ and $S_2$ in the Raman spectrum and the HR-S/TEM image we conclude that the Type 3 sample is more likely to be the A′B′B stacking sequence. There is a possibility that the Type 3 actually corresponds to the ABA stacking sequence (Type 4 Raman spectrum), but the forbidden Raman peak at 9.6 cm$^{-1}$ appears due to disorder or other effects that relax the selection rule. However, we found that there is bilayer region with the AA′-type stacking in the surrounding area, which supports our assignment that this region is indeed the A′B′B stacking sequence. Finally, the identification of Type 1 is more challenging: from the Raman analysis, it was assigned to either AA′B′ or A′B′B stacking sequences. However, the HR-TEM analysis indicates that it is the ABC stacking sequence although some disorder was observed in the surrounding region (see Fig. S8†). We interpret that this region is indeed the ABC stacking sequence, but the forbidden Raman peak of $S_2$ (16.6 cm$^{-1}$) appears due to the relaxation of the selection rule caused by the disorder because the laser spot for the Raman measurement is much larger than the sampling area of the HR-S/TEM analysis.

We never found the AA′A stacking sequence in our HR-S/TEM measurements, although this stacking sequence (β-GaSe in bulk) has been reported in bulk studies.[53] Furthermore, in many ultra-low-frequency Raman spectra that we measured, we never encountered a spectrum with a different position for the shear mode $S_2$ as expected from the AA′A stacking sequence according to our calculations (Fig. 4a). We suspect that this stacking sequence does not exfoliate as easily as the other stacking sequences or this stacking is unstable in few-layers or it could easily transit to other stacking sequences. The classification of stacking sequences is summarized in table 1. Among many samples that we measured, the Type 2 (ABC) and the Type 4 (ABA) spectra were more frequently observed than the other types, which implies that the AB-type stacking is probably more stable and common than the AA′-type stacking. This is corroborated by our theoretical calculations: for bilayer, the AB stacking is energetically more stable than the AA′ stacking by ~1.1 meV per unit cell; for trilayer, the ABA, ABC, and A′B′B stacking patterns are more stable than the AA'A stacking by 2.3 meV per unit cell, 2.4 meV per unit cell, and 1.2 meV per unit cell, respectively. These trends do suggest that the AB stacking between two adjacent layers should be more common than the AA' stacking.

## 4. CONCLUSIONS

We investigated 2D layered GaSe by using polarized Raman spectroscopy as a function of the number of layers and the polarization angle. We found a blueshift of the high-frequency intra-layer $A_1'(1)$ modes as the number of layers increases whereas other high-frequency peak positions do not change much. More importantly, the low-frequency shear and breathing modes from inter-layer vibrations were observed, and they are more sensitive to the thickness and stacking. We can determine the number of layers with the peak positions of the low-frequency shear modes and the mode along with AFM. We found that the ultra-low-frequency spectra are much different even for the same thickness in the same flake. By comparing the ultra-low-frequency Raman spectra with theoretical calculations and HR-S/TEM measurements, we established the correlation between the ultra-low-frequency Raman spectra and the stacking sequences for trilayer GaSe. We further found that the AB-type stacking is more stable than the AA′-type stacking in GaSe. Our findings demonstrate that the inter-layer shear modes (frequencies and intensities) can be effective indicators of thickness and stacking in few-layer GaSe, the two key parameters governing the electronic and optical properties of 2D materials.

**Author contribution**

S.Y.L. prepared samples and carried out the Raman measurements. T.T.H.N and S.C. provided the bulk crystal. L.L. and X.K. calculated theoretical Raman spectra and contributed to theoretical analysis. J.H.K. and Z.L. carried out the S/TEM analysis. S.Y.L., J.-U.L., L.L., and H.C. interpreted the spectroscopic data.

**Conflicts of interest**

There are no conflicts to declare.

**Acknowledgements**

This work was supported by the National Research Foundation (NRF) grant funded by the Korean government (MSIT)(NRF-2014R1A4A1071686, 2018R1A2A2A05019598, 2019R1A2C3006189, 2019R1F1A1058473, and No. 2017R1A5A1014862, SRC program: vdWMRC center), by a grant IBS-R019-D1, and by a grant (No. 2013M3A6A5073173) from the Center for Advanced Soft Electronics under the Global Frontier Research Program of MSIT. A portion of this research (Raman scattering modeling) used resources at the Center for Nanophase Materials Sciences, which is a US Department of Energy Office of Science User Facility. L.L.

and X.K. acknowledge work conducted at the Center for Nanophase Materials Sciences. S.Y.L. acknowledges support from Hyundai Motor Chung Mong-Koo Foundation.

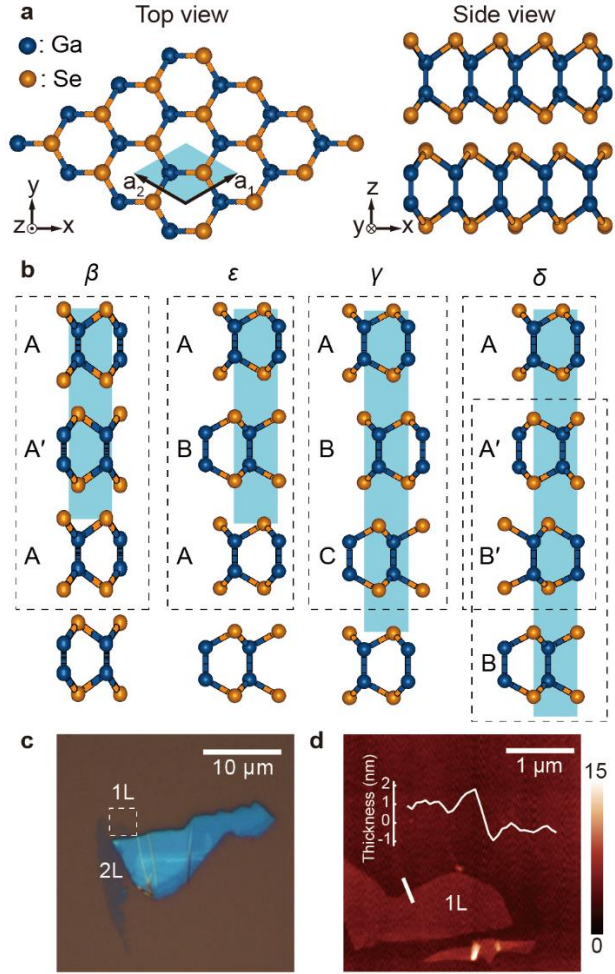

**Fig. 1** (a) Top and side view of GaSe crystal. The unit cell is indicated with a light-blue box. (b) Four different polytypes for bulk GaSe: $\beta$, $\varepsilon$, $\gamma$, and $\delta$-type. The light-blue boxes are the bulk unit cells for each polytype, and the dashed boxes indicate the stacking sequences of trilayer GaSe. (c) Optical image of a typical GaSe flake on SiO$_2$/Si substrate. The white dashed box is the region where the AFM measurement was performed. (d) AFM image of the area indicated in (c). The line scan confirms the thickness of the monolayer region.

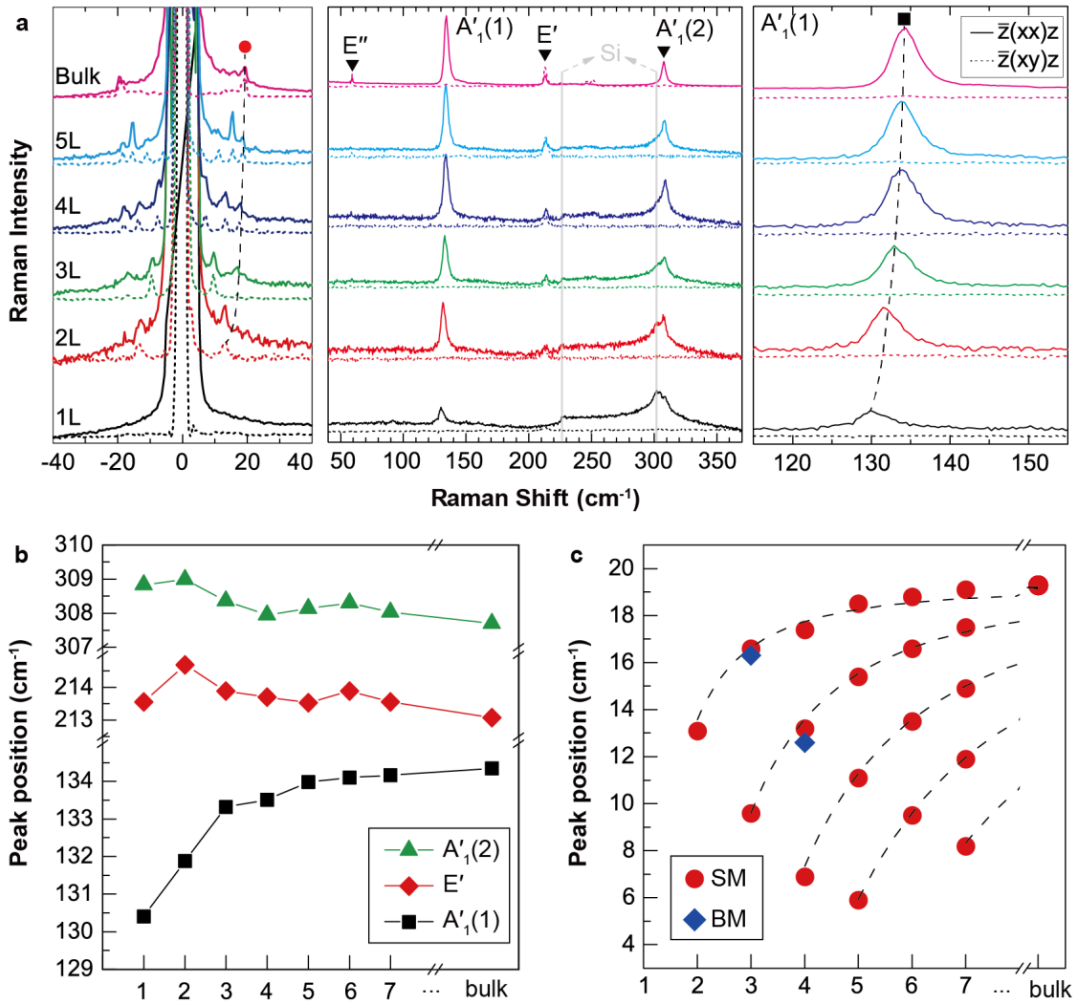

**Fig. 2** (a) Representative Raman spectra of GaSe samples from monolayer to bulk measured with a 532-nm (2.33-eV) excitation energy. The solid gray lines indicate the signal from Si substrates. The magnified ultra-low-frequency spectra are shown in Fig. S2†. (b) Thickness dependence of the high-frequency intra-layer vibrational modes: $A'_1(1)$, $E'$, and $A'_1(2)$. (c) Measured peak positions of the inter-layer shear modes (SM) and the inter-layer breathing modes (BM) as a function of the number of layers. The dashed curves are the best fit to linear chain model calculations.

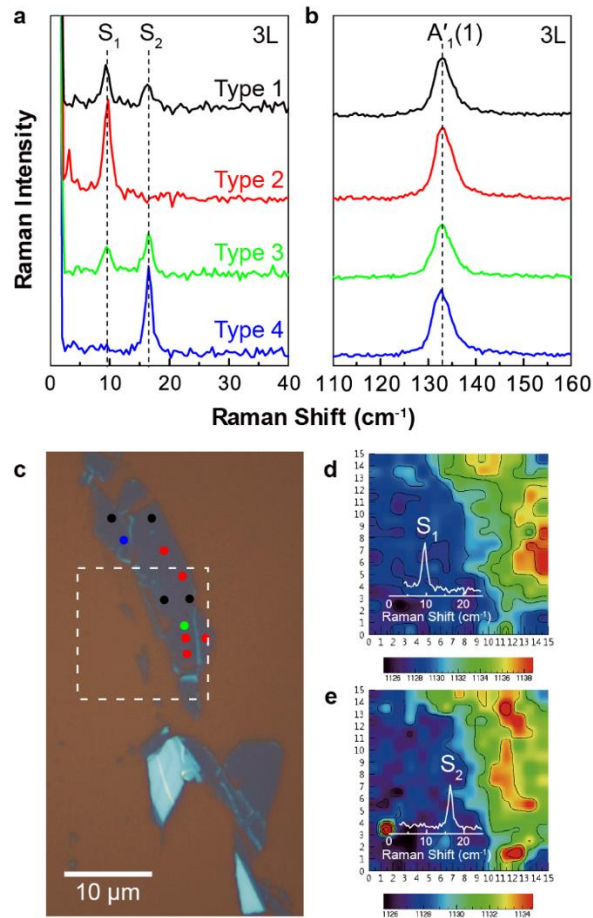

**Fig. 3** (a) Four different types of ultra-low-frequency Raman spectra measured in cross-polarization configuration for trilayer GaSe. (b) The $A'_1(1)$ mode for each stacking sequence measured in parallel polarization configuration. (c) Optical image of a trilayer GaSe sample where multiple positions were measured. The colored circles indicate the type of the ultra-low-frequency Raman spectrum: black, red, green, and blue circles correspond to the color of the spectra in (a) and (b). (d,e) Raman intensity maps taken from the area indicated by the dashed box in (c), corresponding to peaks (d) $S_1$ and (e) $S_2$ in the ultra-low-frequency Raman spectra of (a).

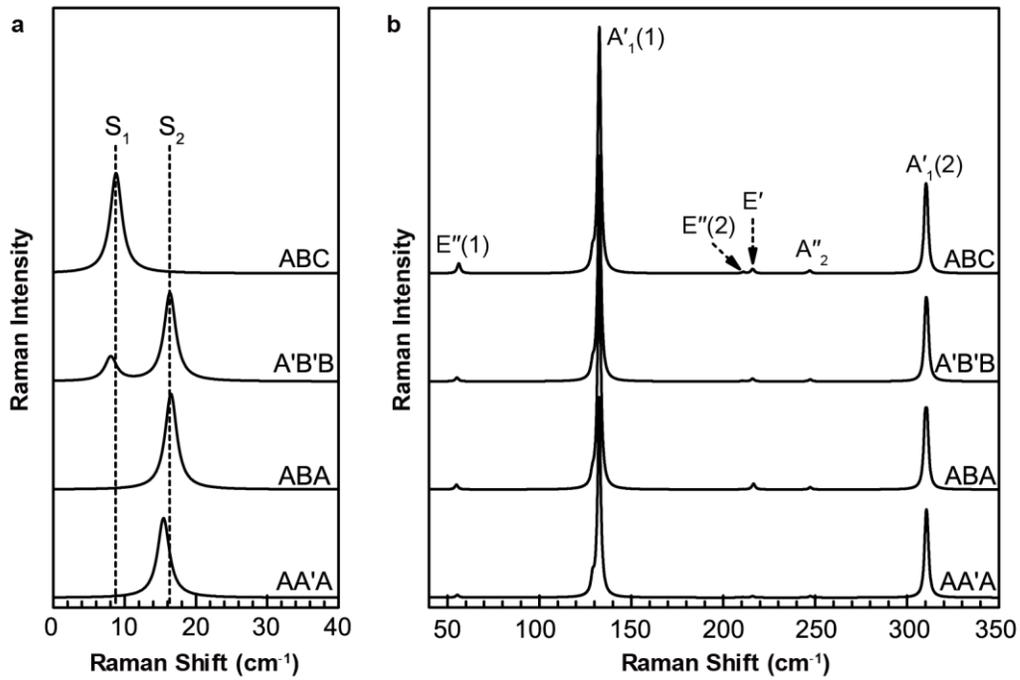

**Fig. 4** Calculated non-resonant Raman spectra for trilayer GaSe with respect to different stacking sequences based on DFT. (a) Ultra-low-frequency Raman spectra for trilayer GaSe which show inter-layer shear modes $S_1$ and $S_2$. Note that the $S_1$ peak intensity in A′B′B stacking was increased by one order of magnitude for better display. (b) High-frequency Raman spectra which show intra-layer vibrational modes $E''(1)$, $A'_1(1)$, $E''(2)$, $E'$, $A''_2$ and $A'_1(2)$, following the bulk symmetry notations.

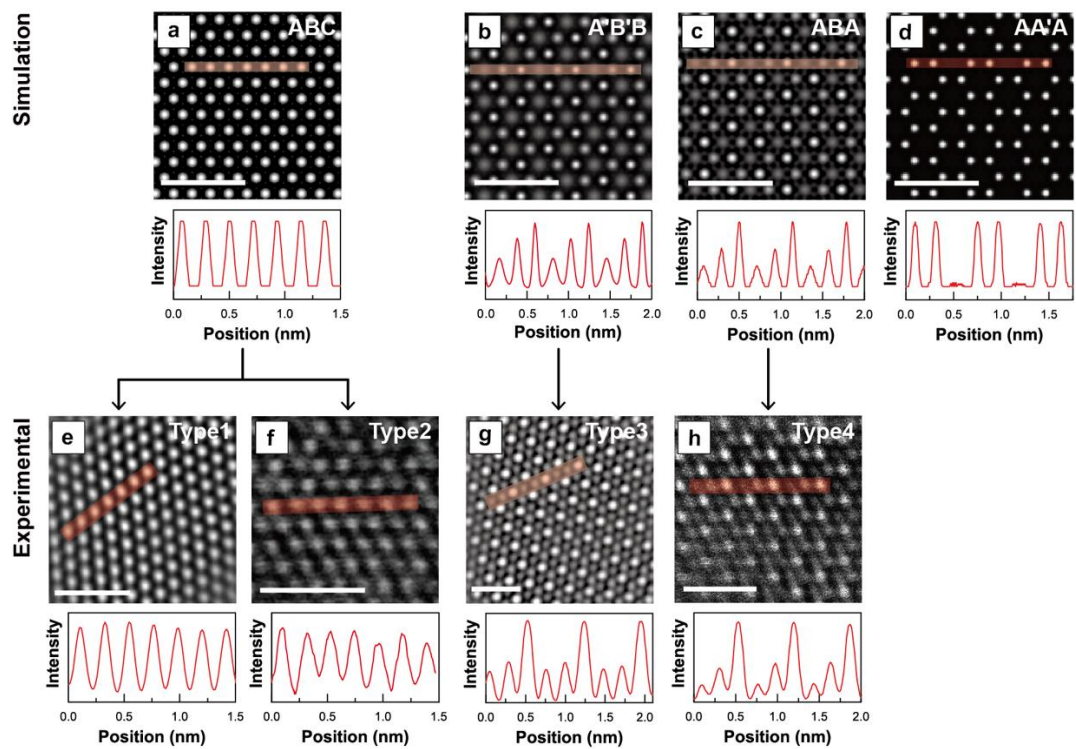

**Fig. 5** HR-STEM and HR-TEM results from four different types of trilayer samples. (a-d) Simulated HR-TEM images of trilayer GaSe for the stacking sequences indicated. (e) Measured HR-TEM image and intensity line profile along the red line in the image for Type 1 sample. (f-h) The HR-STEM images and intensity line profiles along the red lines in each high-resolution image for Type 2, Type 3, and Type 4, respectively. The scale bar is 1 nm.

**Table 1.** Assignment of stacking sequences of trilayer GaSe samples

| Raman Spectrum | Theory and HR-S/TEM |
|---|---|
| Type 1 | Disordered ABC |
| Type 2 | ABC |
| Type 3 | A′B′B |
| Type 4 | ABA |

# Supplementary Information

# Polytypism in Few-Layer Gallium Selenide

Soo Yeon Lim[a], Jae-Ung Lee[a,b], Jung Hwa Kim[c,d], Liangbo Liang[e], Xiangru Kong[e], Thi Thanh Huong Nguyen[f], Zonghoon Lee[c,d], Sunglae Cho[f] and Hyeonsik Cheong[a]

[a] Department of Physics, Sogang University, Seoul 04107, Korea
[b] Department of Physics, Ajou university, Suwon 16499, Korea
[c] School of Materials Science and Engineering, Ulsan National Institute of Science and Technology (UNIST), Ulsan 44919, Korea
[d] Center for Multidimensional Carbon Materials, Institute for Basic Science (IBS), Ulsan 44919, Korea
[e] Center for Nanophase Materials Sciences, Oak Ridge National Laboratory, Oak Ridge, Tennessee 37831, United States
[f] Department of Physics and Energy Harvest Storage Research Center, University of Ulsan, Ulsan 44610, Korea

E-mail: hcheong@sogang.ac.kr

**Contents:**

- **Fig. S1.** Optical and AFM images showing laser-induced degradation of few-layer GaSe

- **Fig. S2.** Ultra-low-frequency Raman spectra for 1-, 2-, 3-, 4-, 5-layer, and bulk GaSe in cross and parallel polarization configurations.

- **Fig. S3.** Polarization dependence of Raman peak intensities of (a) $A'_1(1)$, (b) $E'$, and (c) $A'_1(2)$ modes measured in the parallel polarization configuration, showing in-plane isotropy.

- **Fig. S4.** Raman spectra of 6- and 7-layer and bulk GaSe.

- **Note S1.** Raman tensor calculations for four polytypes of bulk GaSe in linear and circular polarization configurations.

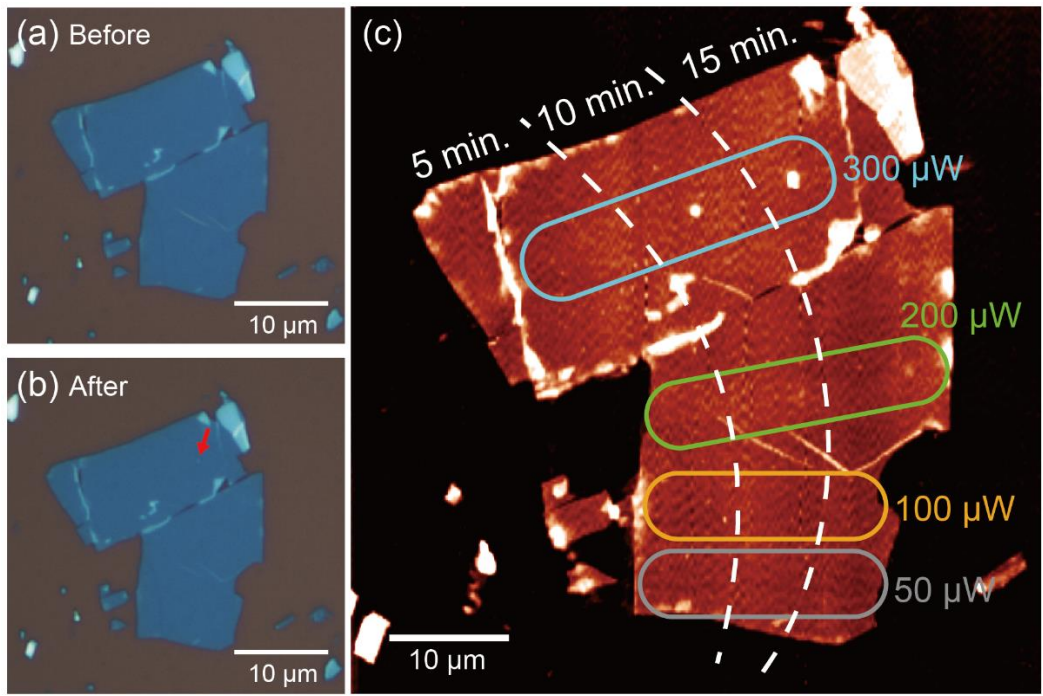

**Fig. S1.** Laser-induced degradation of few-layer GaSe by a 532-nm laser. Optical images taken (a) before and (b) after laser exposure in vacuum showing that damages are not visible except for the case of 15-min irradiation with the power of 300 µW. (c) AFM image of the sample after laser exposure. The laser powers and the exposure times are indicated. For 100 µW or below, no apparent change is observed. At 200 µW, a slight change is seen after 10 min. At 300 µW, obvious degradation is seen after 10 min.

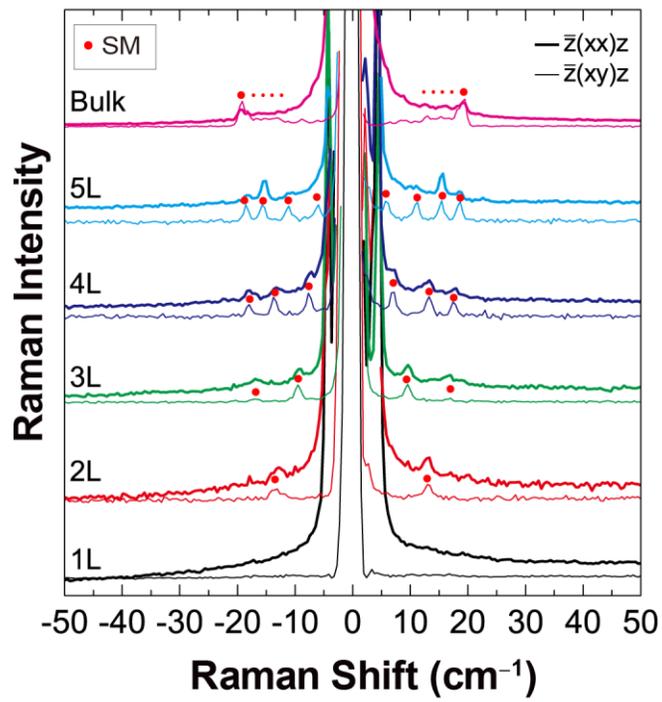

**Fig. S2**. Ultra-low-frequency Raman spectra for 1-, 2-, 3-, 4-, 5-layer, and bulk GaSe in cross and parallel polarization configurations.

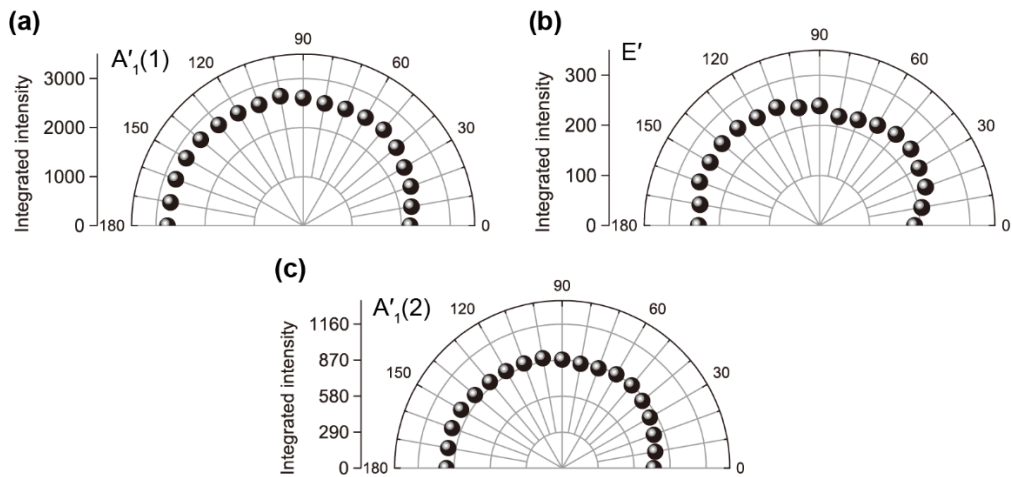

**Fig. S3.** Polarization dependence of Raman peak intensities of (a) $A_1'(2)$, (b) $E'$, and (c) $A_1'(2)$ modes measured in the parallel polarization configuration, showing in-plane isotropy.

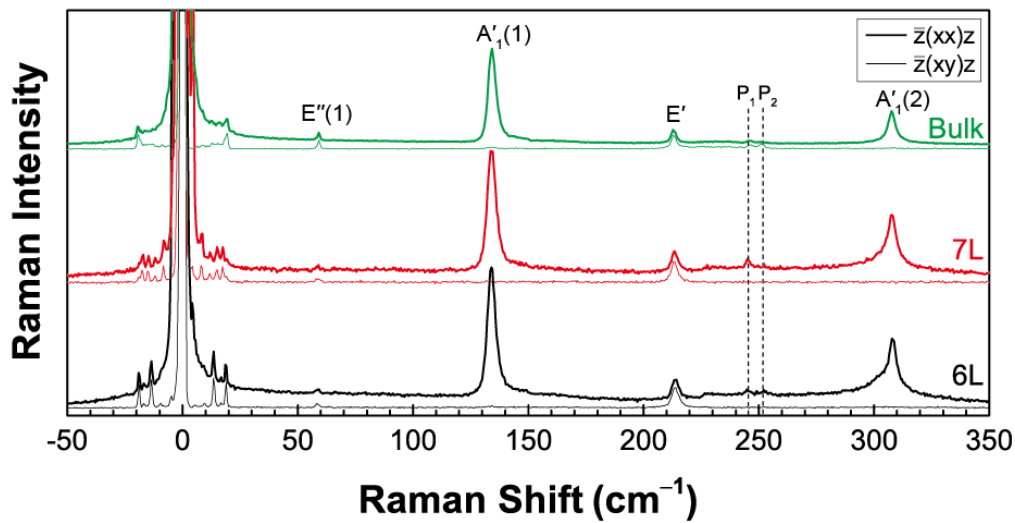

**Fig. S4.** Raman spectra of 6- and 7-layer and bulk GaSe. $P_1$ and $P_2$ are occasionally observed in thick samples.

**Note S1. Raman tensor calculations for four polytypes of bulk GaSe in linear and circular polarization configurations**[S1, S2]

The *β*- and *ε*-GaSe have 24 normal vibrational modes with $D_{6h}^4$ and $D_{3h}^1$ space group, respectively. Since the *γ*-GaSe with $C_{3v}^5$ space group includes three layers in one unit cell, 36 normal modes exist. The *δ*-GaSe has $C_{6v}^4$ space group corresponding to 48 normal modes. The four different polytypes of bulk GaSe show the following irreducible representations at the zone center:

$$\Gamma_\beta = 2A_{2u} + 2A_{1g} + 2B_{1u} + 2B_{2g} + 2E_{1u} + 2E_{2g} + 2E_{1g} + 2E_{2u}$$

$$\Gamma_\varepsilon = 4A_1' + 4A_2'' + 4E' + 4E''$$

$$\Gamma_\gamma = 12A_1 + 12E$$

$$\Gamma_\delta = 8A_1 + 8B_1 + 8E_1 + 8E_2$$

(S1)

The *β*-GaSe has 6 Raman active modes ($2A_{1g}$, $2E_{2g}$, and $2E_{1g}$ modes) and the *ε*-GaSe have 11 nondegenerate Raman active modes ($4A_1'$, $3E'$, and $4E_1''$). On the other hand, all the optical modes in *γ*-GaSe are both infrared and Raman active, so 22 nondegenerate Raman active modes exist. For the *δ*-GaSe, 7 $A_1$ and 7 $E_1$ modes except the acoustic modes are Raman allowed as well as 8 modes of $E_2$.

The intensity of each mode is proportional to $|\hat{e}_s \cdot \overset{=}{R} \cdot \hat{e}_i|^2$, where $\hat{e}_s$ and $\hat{e}_i$ are polarizations of the scattered and incident photons, respectively, and $\overset{=}{R}$ is the Raman tensor. For the hexagonal (*β*-, *ε*-, and *δ*-) GaSe, the Raman tensors can be written as

$$\beta\text{-GaSe} : A_{1g} = \begin{pmatrix} a & 0 & 0 \\ 0 & a & 0 \\ 0 & 0 & b \end{pmatrix}, E_{1g} = \begin{pmatrix} 0 & 0 & 0 \\ 0 & 0 & c \\ 0 & c & 0 \end{pmatrix}, E_{1g} = \begin{pmatrix} 0 & 0 & -c \\ 0 & 0 & 0 \\ -c & 0 & 0 \end{pmatrix},$$

$$E_{2g} = \begin{pmatrix} d & 0 & 0 \\ 0 & -d & 0 \\ 0 & 0 & 0 \end{pmatrix}, E_{2g} = \begin{pmatrix} 0 & -d & 0 \\ -d & 0 & 0 \\ 0 & 0 & 0 \end{pmatrix}$$

(S2)

$$\varepsilon\text{-GaSe} : A'_1 = \begin{pmatrix} a & 0 & 0 \\ 0 & a & 0 \\ 0 & 0 & b \end{pmatrix}, E'' = \begin{pmatrix} 0 & 0 & 0 \\ 0 & 0 & c \\ 0 & c & 0 \end{pmatrix}, E'' = \begin{pmatrix} 0 & 0 & -c \\ 0 & 0 & 0 \\ -c & 0 & 0 \end{pmatrix},$$

$$E'(x) = \begin{pmatrix} d & 0 & 0 \\ 0 & -d & 0 \\ 0 & 0 & 0 \end{pmatrix}, E'(y) = \begin{pmatrix} 0 & -d & 0 \\ -d & 0 & 0 \\ 0 & 0 & 0 \end{pmatrix}$$

(S3)

$$\delta\text{-GaSe} : A_1(z) = \begin{pmatrix} a & 0 & 0 \\ 0 & a & 0 \\ 0 & 0 & b \end{pmatrix}, E_1(x) = \begin{pmatrix} 0 & 0 & c \\ 0 & 0 & 0 \\ c & 0 & 0 \end{pmatrix}, E_1(y) = \begin{pmatrix} 0 & 0 & 0 \\ 0 & 0 & c \\ 0 & c & 0 \end{pmatrix},$$

$$E_2 = \begin{pmatrix} d & 0 & 0 \\ 0 & -d & 0 \\ 0 & 0 & 0 \end{pmatrix}, E_2 = \begin{pmatrix} 0 & -d & 0 \\ -d & 0 & 0 \\ 0 & 0 & 0 \end{pmatrix}$$

(S4)

where $a$, $b$, $c$ and $d$ are constants. On the other hand, $\gamma$-GaSe has the Raman tensors:

$$\gamma\text{-GaSe} : A_1(z) = \begin{pmatrix} a & 0 & 0 \\ 0 & a & 0 \\ 0 & 0 & b \end{pmatrix}, E(x) = \begin{pmatrix} 0 & c & d \\ c & 0 & 0 \\ d & 0 & 0 \end{pmatrix}, E(y) = \begin{pmatrix} c & 0 & 0 \\ 0 & -c & d \\ 0 & d & 0 \end{pmatrix}$$

(S5)

For the case of the linearly polarized light in back-scattering geometry, we can write down the incident and scattered light vectors as $\hat{e}_i = (\cos\theta_i \quad \sin\theta_i \quad 0)$ and $\hat{e}_s = (\cos\theta_s \quad \sin\theta_s \quad 0)$, respectively. The angles $\theta_i$ and $\theta_s$ are angle of the incident and scattered light polarization with respect to an arbitrary reference direction 0°. Therefore, the intensities depending on the polarization angle of $A_{1g}$ modes for $\beta$-GaSe, for example, can be calculated as

$$I_{A_{1g}} \propto \left|\hat{e}_s \cdot \ddot{R} \cdot \hat{e}_i\right|^2 = \left|(\cos\theta_s \quad \sin\theta_s \quad 0)\begin{pmatrix} a & 0 & 0 \\ 0 & a & 0 \\ 0 & 0 & b \end{pmatrix}\begin{pmatrix} \cos\theta_i \\ \sin\theta_i \\ 0 \end{pmatrix}\right|^2 = a^2\left|\cos(\theta_i - \theta_s)\right|^2 \tag{S6}$$

If the incident and scattered light are perpendicular, the term is zero. Therefore, we can only observe the $A_{1g}$ modes in the parallel polarization configuration and not in the cross polarization configuration. Since the $A_1'$ mode for $\varepsilon$-GaSe and $A_1(z)$ modes for $\delta$-GaSe and $\gamma$-GaSe have the identical for of the Raman tensors, the polarization dependences are the same.

In contrast to the A modes, some E modes are forbidden in back-scattering geometry although they are Raman active. The Raman tensors of two $E_{1g}$ modes for $\beta$-GaSe, two of E″ modes for $\varepsilon$-GaSe, and $E_1(x)$ and $E_1(y)$ modes for $\delta$-GaSe have no elements in the first and second columns and rows. Therefore, these E modes are forbidden in the back-scattering geometry. On the other hand, the intensities of two $E_{2g}$ modes for $\beta$-GaSe depending on the polarization angle are

$$I_{E_{2g}} \propto \left|(\cos\theta_s \quad \sin\theta_s \quad 0)\begin{pmatrix} d & 0 & 0 \\ 0 & -d & 0 \\ 0 & 0 & 0 \end{pmatrix}\begin{pmatrix} \cos\theta_i \\ \sin\theta_i \\ 0 \end{pmatrix}\right|^2 = d^2\left|\cos(\theta_i + \theta_s)\right|^2, \tag{S7}$$

$$I_{E_{2g}} \propto \left|(\cos\theta_s \quad \sin\theta_s \quad 0)\begin{pmatrix} 0 & -d & 0 \\ -d & 0 & 0 \\ 0 & 0 & 0 \end{pmatrix}\begin{pmatrix} \cos\theta_i \\ \sin\theta_i \\ 0 \end{pmatrix}\right|^2 = d^2\left|\sin(\theta_i + \theta_s)\right|^2 \tag{S8}$$

Since these two $E_{2g}$ modes are degenerate, the polarization dependence has a superposed form of the two modes, resulting in the total intensity proportional to a constant $d^2$. Therefore, the $E_{2g}$ modes can be observed regardless of the polarization configuration. The $E'(x)$ and $E'(y)$ modes for $\varepsilon$-GaSe and the $E_2$ modes for $\delta$-GaSe have the identical tensor forms, leading to the identical polarization dependences.

The Raman tensors of the $E(x)$ and $E(y)$ modes for the $\gamma$-Gase are $\begin{pmatrix} 0 & c & d \\ c & 0 & 0 \\ d & 0 & 0 \end{pmatrix}$ and $\begin{pmatrix} c & 0 & 0 \\ 0 & -c & d \\ 0 & d & 0 \end{pmatrix}$, respectively. The intensities are

$$I_{E(x)} \propto \left| \begin{pmatrix} \cos\theta_s & \sin\theta_s & 0 \end{pmatrix} \begin{pmatrix} 0 & c & d \\ c & 0 & 0 \\ d & 0 & 0 \end{pmatrix} \begin{pmatrix} \cos\theta_i \\ \sin\theta_i \\ 0 \end{pmatrix} \right|^2 = c^2 \left| \sin(\theta_i + \theta_s) \right|^2 \text{ and} \tag{S9}$$

$$I_{E(y)} \propto \left| \begin{pmatrix} \cos\theta_s & \sin\theta_s & 0 \end{pmatrix} \begin{pmatrix} c & 0 & 0 \\ 0 & -c & d \\ 0 & d & 0 \end{pmatrix} \begin{pmatrix} \cos\theta_i \\ \sin\theta_i \\ 0 \end{pmatrix} \right|^2 = c^2 \left| \cos(\theta_i + \theta_s) \right|^2, \tag{S10}$$

the same as the case of the $E_{2g}$ modes for $\beta$-GaSe: the E modes for $\gamma$-GaSe have no polarization dependence.

Consequently, for linearly polarization configuration, all the A modes are observed in the parallel polarization configuration and not in the cross polarization configuration. The $E_{1g}$ mode for $\beta$-GaSe, $E''$ mode for $\varepsilon$-GaSe, and $E_1(x)$ and $E_1(y)$ modes for $\delta$-GaSe are not allowed in back-scattering geometry. All the other E modes for all the polytypes are allowed and have no polarization dependence. The results are summarized in Table S1.

Circularly polarized light in back-scattering geometry is represented by $\sigma\pm = \frac{1}{\sqrt{2}}(1 \quad \mp i \quad 0)$.

The intensities of $A_{1g}$ modes for all the polytypes are proportional to $a^2$ when the incident and scattered light have the same polarizations $[(\sigma+\sigma+) \text{ or } (\sigma-\sigma-)]$ whereas the intensity is zero with the opposite polarizations $[(\sigma+\sigma-) \text{ or } (\sigma-\sigma+)]$, according to the following calculations:

$$\left|\sigma^\dagger(+)\cdot \ddot{R}\cdot \sigma(+)\right|^2_{A_{1g}} \propto \frac{1}{2}\left|(1 \quad i \quad 0)\begin{pmatrix} a & 0 & 0 \\ 0 & a & 0 \\ 0 & 0 & b \end{pmatrix}\begin{pmatrix} 1 \\ -i \\ 0 \end{pmatrix}\right|^2 = \frac{1}{2}|a+a+0|^2 = 2a^2$$

$$\left|\sigma^\dagger(-)\cdot \ddot{R}\cdot \sigma(-)\right|^2_{A_{1g}} \propto \frac{1}{2}\left|(1 \quad -i \quad 0)\begin{pmatrix} a & 0 & 0 \\ 0 & a & 0 \\ 0 & 0 & b \end{pmatrix}\begin{pmatrix} 1 \\ i \\ 0 \end{pmatrix}\right|^2 = \frac{1}{2}|a+a+0|^2 = 2a^2$$

$$\left|\sigma^\dagger(+)\cdot \ddot{R}\cdot \sigma(-)\right|^2_{A_{1g}} \propto \frac{1}{2}\left|(1 \quad i \quad 0)\begin{pmatrix} a & 0 & 0 \\ 0 & a & 0 \\ 0 & 0 & b \end{pmatrix}\begin{pmatrix} 1 \\ i \\ 0 \end{pmatrix}\right|^2 = \frac{1}{2}|a-a+0|^2 = 0$$

$$\left|\sigma^\dagger(-)\cdot \ddot{R}\cdot \sigma(+)\right|^2_{A_{1g}} \propto \frac{1}{2}\left|(1 \quad -i \quad 0)\begin{pmatrix} a & 0 & 0 \\ 0 & a & 0 \\ 0 & 0 & b \end{pmatrix}\begin{pmatrix} 1 \\ -i \\ 0 \end{pmatrix}\right|^2 = \frac{1}{2}|a-a+0|^2 = 0 \quad (S11)$$

All the other A modes for other polytypes have the same polarization dependences.

As we mentioned above, two $E_{1g}$ modes for $\beta$-GaSe, two of the $E''$ modes for $\varepsilon$-GaSe, and the $E_1(x)$ and $E_1(y)$ modes for $\delta$-GaSe are forbidden in back-scattering geometry. The Raman tensors of the $E_{2g}$ modes for $\beta$-GaSe, the $E'(x)$ and $E'(y)$ modes for $\varepsilon$-GaSe, and the $E_2$ modes for δ-GaSe are $\begin{pmatrix} d & 0 & 0 \\ 0 & -d & 0 \\ 0 & 0 & 0 \end{pmatrix}$ and $\begin{pmatrix} 0 & -d & 0 \\ -d & 0 & 0 \\ 0 & 0 & 0 \end{pmatrix}$. These show opposite tendency compared to the A modes in terms of the polarization dependences.

$$\left| \sigma^\dagger(+) \cdot \ddot{R} \cdot \sigma(+) \right|^2_{E_{2g},\, E'(x),\, E_2} \propto \frac{1}{2} \left| (1\ i\ 0) \begin{pmatrix} d & 0 & 0 \\ 0 & -d & 0 \\ 0 & 0 & 0 \end{pmatrix} \begin{pmatrix} 1 \\ -i \\ 0 \end{pmatrix} \right|^2 = \frac{1}{2} |d - d + 0|^2 = 0$$

$$\left| \sigma^\dagger(-) \cdot \ddot{R} \cdot \sigma(-) \right|^2_{E_{2g},\, E'(x),\, E_2} \propto \frac{1}{2} \left| (1\ -i\ 0) \begin{pmatrix} d & 0 & 0 \\ 0 & -d & 0 \\ 0 & 0 & 0 \end{pmatrix} \begin{pmatrix} 1 \\ i \\ 0 \end{pmatrix} \right|^2 = \frac{1}{2} |d - d + 0|^2 = 0$$

$$\left| \sigma^\dagger(+) \cdot \ddot{R} \cdot \sigma(-) \right|^2_{E_{2g},\, E'(x),\, E_2} \propto \frac{1}{2} \left| (1\ i\ 0) \begin{pmatrix} d & 0 & 0 \\ 0 & -d & 0 \\ 0 & 0 & 0 \end{pmatrix} \begin{pmatrix} 1 \\ i \\ 0 \end{pmatrix} \right|^2 = \frac{1}{2} |d + d + 0|^2 = 2d^2$$

$$\left| \sigma^\dagger(-) \cdot \ddot{R} \cdot \sigma(+) \right|^2_{E_{2g},\, E'(x),\, E_2} \propto \frac{1}{2} \left| (1\ -i\ 0) \begin{pmatrix} d & 0 & 0 \\ 0 & -d & 0 \\ 0 & 0 & 0 \end{pmatrix} \begin{pmatrix} 1 \\ -i \\ 0 \end{pmatrix} \right|^2 = \frac{1}{2} |d + d + 0|^2 = 2d^2 \quad \text{(S12)}$$

Similarly, for $\begin{pmatrix} 0 & -d & 0 \\ -d & 0 & 0 \\ 0 & 0 & 0 \end{pmatrix}$, the results are exactly same.

The Raman tensors of $E(x) = \begin{pmatrix} 0 & c & d \\ c & 0 & 0 \\ d & 0 & 0 \end{pmatrix}$, $E(y) = \begin{pmatrix} c & 0 & 0 \\ 0 & -c & d \\ 0 & d & 0 \end{pmatrix}$ for δ-GaSe result in the identical dependence. For the $E(x)$ and $E(y)$ mode, the intensity is zero in the same polarization configuration and proportional to $c^2$ in the opposite polarization configuration.

$$\left|\sigma^{\dagger}(+)\cdot\ddot{\mathrm{R}}\cdot\sigma(+)\right|^{2}_{\mathrm{E}(x)} \propto \frac{1}{2}\left|(1 \ i \ 0)\begin{pmatrix} 0 & c & d \\ c & 0 & 0 \\ d & 0 & 0 \end{pmatrix}\begin{pmatrix} 1 \\ -i \\ 0 \end{pmatrix}\right|^{2} = \frac{1}{2}\left|ic-ic+0\right|^{2}=0$$

$$\left|\sigma^{\dagger}(-)\cdot\ddot{\mathrm{R}}\cdot\sigma(-)\right|^{2}_{\mathrm{E}(x)} \propto \frac{1}{2}\left|(1 \ -i \ 0)\begin{pmatrix} 0 & c & d \\ c & 0 & 0 \\ d & 0 & 0 \end{pmatrix}\begin{pmatrix} 1 \\ i \\ 0 \end{pmatrix}\right|^{2} = \frac{1}{2}\left|-ic+ic+0\right|^{2}=0$$

$$\left|\sigma^{\dagger}(+)\cdot\ddot{\mathrm{R}}\cdot\sigma(-)\right|^{2}_{\mathrm{E}(x)} \propto \frac{1}{2}\left|(1 \ i \ 0)\begin{pmatrix} 0 & c & d \\ c & 0 & 0 \\ d & 0 & 0 \end{pmatrix}\begin{pmatrix} 1 \\ i \\ 0 \end{pmatrix}\right|^{2} = \frac{1}{2}\left|2ic\right|^{2}=2c^{2}$$

$$\left|\sigma^{\dagger}(-)\cdot\ddot{\mathrm{R}}\cdot\sigma(+)\right|^{2}_{\mathrm{E}(x)} \propto \frac{1}{2}\left|(1 \ -i \ 0)\begin{pmatrix} 0 & c & d \\ c & 0 & 0 \\ d & 0 & 0 \end{pmatrix}\begin{pmatrix} 1 \\ -i \\ 0 \end{pmatrix}\right|^{2} = \frac{1}{2}\left|-2ic\right|^{2}=2c^{2} \qquad (S13)$$

Consequently, the A modes are observed in the same polarization configuration and the E modes in the opposite polarization configuration. It is possible to differentiate the A and E modes by using the circularly polarized light. The results are summarized in Table S1.

**Table S1.** Raman intensities of bulk GaSe in linear and circular polarization configurations

| Polytype | Vibrational modes | $\bar{z}(xx)z$ | $\bar{z}(xy)z$ | $\sigma\pm\sigma\pm$ | $\sigma\pm\sigma\mp$ |
|---|---|---|---|---|---|
| $\beta$ | $A_{1g} = \begin{pmatrix} a & 0 & 0 \\ 0 & a & 0 \\ 0 & 0 & b \end{pmatrix}$ | $\propto a^2$ | 0 | $\propto a^2$ | 0 |
| | $E_{1g} = \begin{pmatrix} 0 & 0 & 0 \\ 0 & 0 & c \\ 0 & c & 0 \end{pmatrix}$ | 0 | 0 | 0 | 0 |
| | $E_{1g} = \begin{pmatrix} 0 & 0 & -c \\ 0 & 0 & 0 \\ -c & 0 & 0 \end{pmatrix}$ | | | | |
| | $E_{2g} = \begin{pmatrix} d & 0 & 0 \\ 0 & -d & 0 \\ 0 & 0 & 0 \end{pmatrix}$ | $\propto d^2$ | $\propto d^2$ | 0 | $\propto d^2$ |
| | $E_{2g} = \begin{pmatrix} 0 & -d & 0 \\ -d & 0 & 0 \\ 0 & 0 & 0 \end{pmatrix}$ | | | | |
| $\varepsilon$ | $A_1' = \begin{pmatrix} a & 0 & 0 \\ 0 & a & 0 \\ 0 & 0 & b \end{pmatrix}$ | $\propto a^2$ | 0 | $\propto a^2$ | 0 |
| | $E'' = \begin{pmatrix} 0 & 0 & 0 \\ 0 & 0 & c \\ 0 & c & 0 \end{pmatrix}$ | 0 | 0 | 0 | 0 |
| | $E'' = \begin{pmatrix} 0 & 0 & -c \\ 0 & 0 & 0 \\ -c & 0 & 0 \end{pmatrix}$ | | | | |
| | $E'(x) = \begin{pmatrix} d & 0 & 0 \\ 0 & -d & 0 \\ 0 & 0 & 0 \end{pmatrix}$ | $\propto d^2$ | $\propto d^2$ | 0 | $\propto d^2$ |
| | $E'(y) = \begin{pmatrix} 0 & -d & 0 \\ -d & 0 & 0 \\ 0 & 0 & 0 \end{pmatrix}$ | | | | |
| $\delta$ | $A_1(z) = \begin{pmatrix} a & 0 & 0 \\ 0 & a & 0 \\ 0 & 0 & b \end{pmatrix}$ | $\propto a^2$ | 0 | $\propto a^2$ | 0 |
| | $E_1(x) = \begin{pmatrix} 0 & 0 & c \\ 0 & 0 & 0 \\ c & 0 & 0 \end{pmatrix}$ | 0 | 0 | 0 | 0 |
| | $E_1(y) = \begin{pmatrix} 0 & 0 & 0 \\ 0 & 0 & c \\ 0 & c & 0 \end{pmatrix}$ | | | | |
| | $E_2 = \begin{pmatrix} d & 0 & 0 \\ 0 & -d & 0 \\ 0 & 0 & 0 \end{pmatrix}$ | $\propto d^2$ | $\propto d^2$ | 0 | $\propto d^2$ |
| | $E_2 = \begin{pmatrix} 0 & -d & 0 \\ -d & 0 & 0 \\ 0 & 0 & 0 \end{pmatrix}$ | | | | |
| $\gamma$ | $A_1(z) = \begin{pmatrix} a & 0 & 0 \\ 0 & a & 0 \\ 0 & 0 & b \end{pmatrix}$ | $\propto a^2$ | 0 | $\propto a^2$ | 0 |
| | $E(x) = \begin{pmatrix} 0 & c & d \\ c & 0 & 0 \\ d & 0 & 0 \end{pmatrix}$ | $\propto c^2$ | $\propto c^2$ | 0 | $\propto c^2$ |
| | $E(y) = \begin{pmatrix} c & 0 & 0 \\ 0 & -c & d \\ 0 & d & 0 \end{pmatrix}$ | | | | |

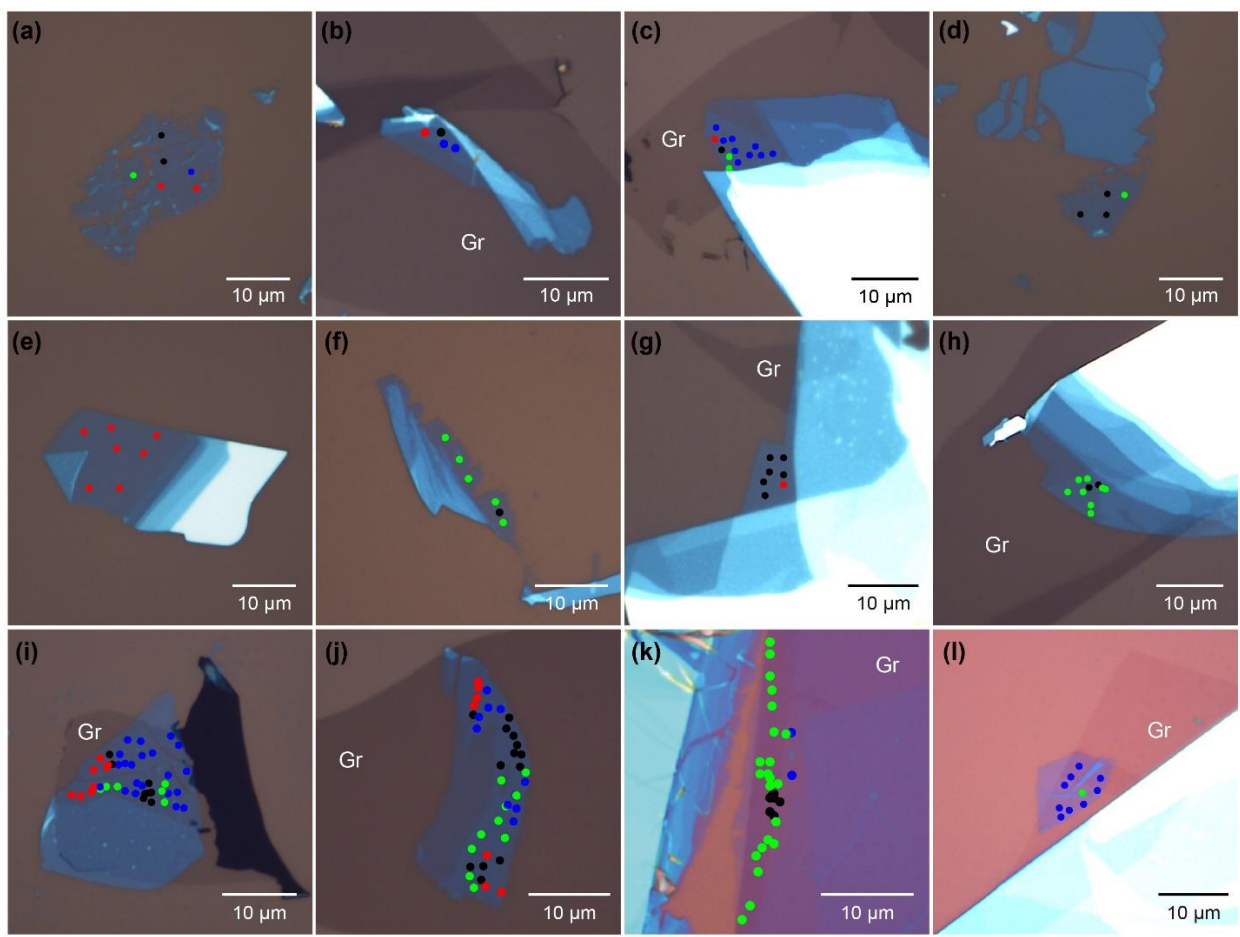

**Fig. S5.** Various trilayer GaSe samples measured in our work. The colored circles indicate the spots where Raman spectra were taken, with the color corresponding to the type of the ultra-low-frequency Raman spectrum: Type 1 (black), Type 2 (red), Type 3 (green), and Type 4 (blue).

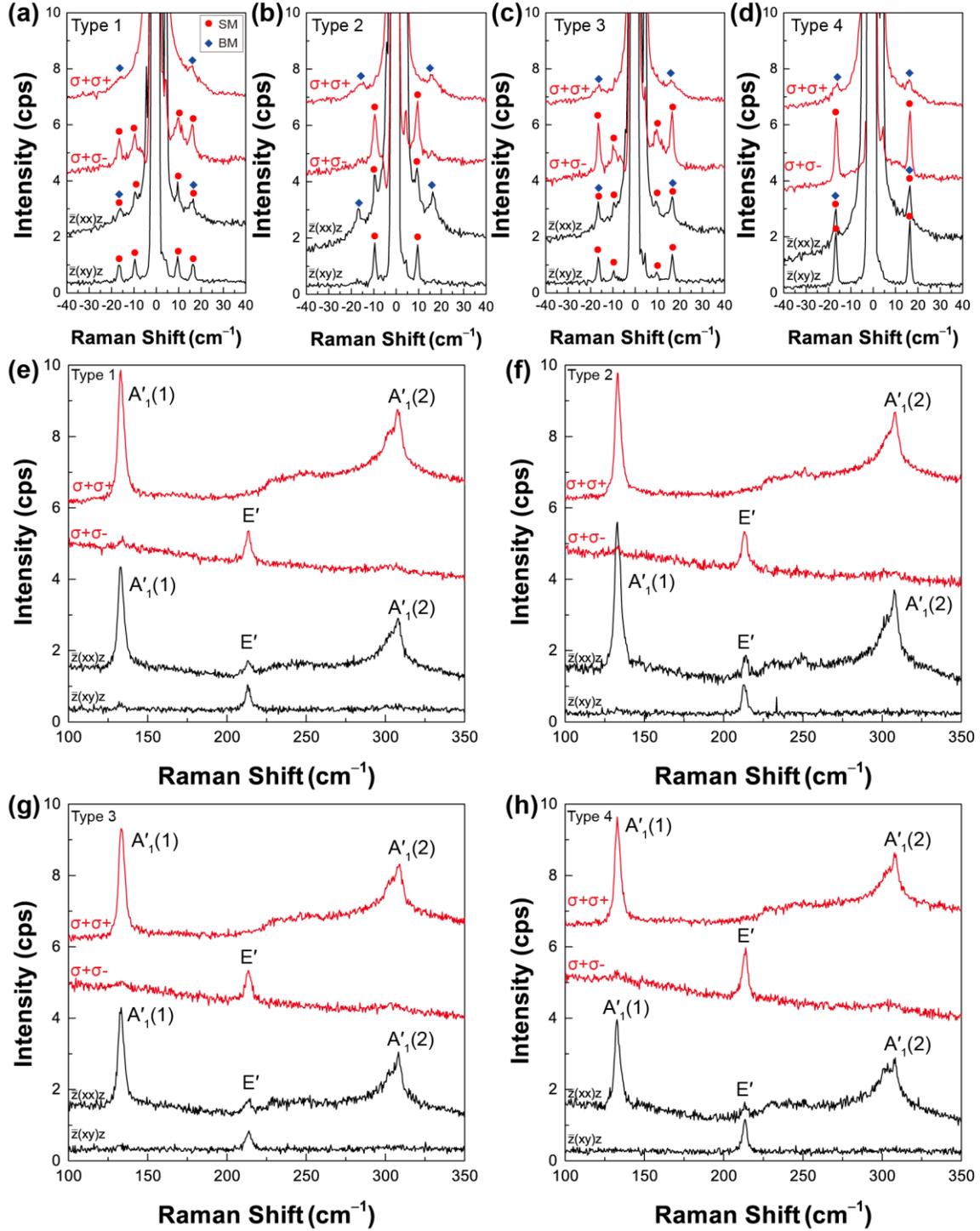

**Fig. S6.** Circularly polarized Raman spectra from trilayer GaSe samples with four different types of the ultra-low-frequency Raman spectra: (a), (e) Type 1, (b), (f) Type 2, (c), (g) Type 3, and (d), (h) Type 4.

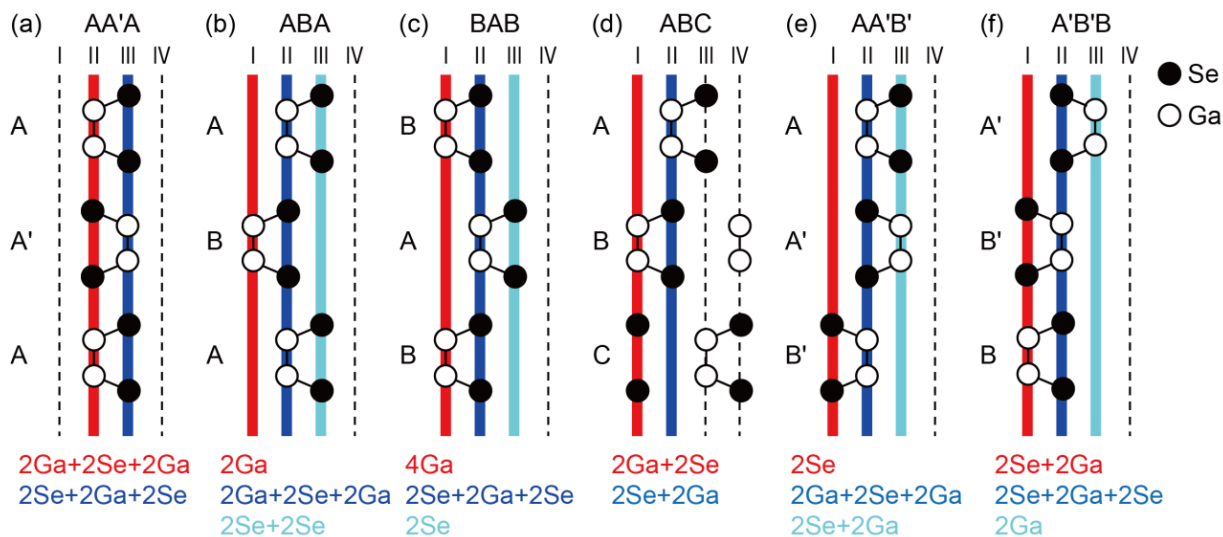

**Fig. S7.** Possible stacking sequences in trilayer GaSe. Each column consists of different atoms in the (a) AA′A, (b) ABA, (c) BAB, (e) AA′B′, and (f) A′B′B, stacking sequences, whereas (d) all the columns in the ABC stacking sequences have the same atoms, two Ga and two Se, which is consistent with S/TEM results.

**Note S2. Inter-layer bond polarizability model**[S3]

In addition to the first-principles DFT method, Raman intensities of low-frequency inter-layer modes in 2D materials can also be computed by a simple inter-layer bond polarizability model.[S3] This model can provide more physical insights compared to the DFT approach. Generally speaking, Raman intensity of each normal mode is proportional to the change of the system's polarizability with respect to the normal coordinates of the corresponding vibration, and so obtaining the polarizability change by the vibration is crucial for calculating the intensity. For an inter-layer vibration mode, each layer oscillates as a quasi-rigid body, and therefore it can be treated as a single object. For the layer $i$, if the derivative of the system's polarizability with respect to its displacement is $\alpha_i'$ and its displacement during the inter-layer vibration is $\Delta r_i$, the change of the polarizability by this displacement is $\Delta \alpha_i = \alpha_i' \cdot \Delta r_i$. The total change of the system's polarizability by the inter-layer vibration is the sum of the changes of every layer: $\Delta \alpha = \sum_i \Delta \alpha_i = \sum_i \alpha_i' \cdot \Delta r_i$, where $\alpha_i'$ is related to the properties of the inter-layer bonds, including the inter-layer bond polarizabilities and the inter-layer bond vectors (lengths and directions).[3] We note that if every layer moves in the same manner (i.e., $\Delta r_i = \Delta r$ for any layer $i$), the polarizability change of the system is given by $\Delta \alpha = (\sum_i \alpha_i') \cdot \Delta r$. Such a motion corresponds to the translation of the whole system by $\Delta r$, and the translational invariance of the system's polarizability requires $\Delta \alpha = 0$, leading to a general relationship $\sum_i \alpha_i' = 0$. For a trilayer system, we then have $\alpha_1' + \alpha_2' + \alpha_3' = 0$. The general form of $\alpha_i'$ can be simply determined based on the directions of the inter-layer bond vectors.[S3,S4] Meanwhile, the displacement of each layer, $\Delta r_i$, can be determined by the linear chain model.[S5] Finally, Raman intensity of the inter-layer vibration mode is obtained based on the formula $I \propto \frac{n+1}{\omega}|\Delta\alpha|^2$, where $n = (e^{\frac{\hbar\omega}{k_B T}} - 1)^{-1}$ is the phonon occupation according to Bose–Einstein statistics and ω is the frequency of the vibration mode.

In trilayer GaSe, for an inter-layer shear vibration along the $x$ direction, the polarizability change is $\Delta\alpha = \sum_i \alpha'_i \cdot \Delta x_i$, where $\alpha'_i$ can change notably with the stacking, since it is sensitive to the inter-layer bond polarizabilities and bond directions that vary with the stacking, according to the inter-layer bond polarizability model.[S3, S4] Bilayer GaSe has two stacking patterns of AA′ and AB, and the inter-layer bond properties are different as the relative layer-layer atomic alignments are different between AA′ and AB stackings. For trilayer GaSe, there are a variety of stacking configurations, including AA′A, ABA, ABC, AA′B′, and A′B′B. Note that both AA′B′ and A′B′B originate from the bulk stacking AA′B′B. For AA′A stacking in trilayer GaSe, the top layer and bottom layer (i.e., layer 1 and layer 3) are in the equivalent positions, thereby giving $\alpha'_1 = \alpha'_3 = \beta'_1$ and subsequently $\alpha'_2 = -2\beta_1$ (recalling the aforementioned general relation $\alpha'_1 + \alpha'_2 + \alpha'_3 = 0$); for ABA stacking, the top layer and bottom layer are also in the equivalent positions, and it has the same form of inter-layer bond vectors as AA′A stacking but different inter-layer bond polarizabilities, therefore giving $\alpha'_1 = \alpha'_3 = \beta_2$ and subsequently $\alpha'_2 = -2\beta_2$ (note that $\beta_1$ and $\beta_2$ are related to the inter-layer bond polarizabilities of AA′ and AB stackings, respectively); for ABC stacking, the layer-layer stacking assumes the same AB type as ABA stacking (i.e., BC stacking equivalent to BA), but layer 2 and layer 3 have different stacking directions and thus the opposite inter-layer bond directions compared to ABA stacking, thus yielding $\alpha'_1 = -\alpha'_3 = \beta_2$ and subsequently $\alpha'_2 = 0$; for AA′B′ stacking, the situation is more complicated due to a mixture of AA′ stacking between layer 1 and layer 2 and AB stacking between layer 2 and layer 3 (A′B′ stacking equivalent to AB), and we can derive that $(\alpha'_1, \alpha'_2, \alpha'_3) = (\beta_1, -\beta_1 + \beta_2, -\beta_2)$; for A′B′B stacking, it is a mixture of AB stacking between layer 1 and layer 2 and A′A stacking between layer 2 and layer 3 (B′B stacking equivalent to A′A, the reversed AA′ stacking), and we can derive that $(\alpha'_1, \alpha'_2, \alpha'_3) = (\beta_2, -\beta_2 - \beta_1, \beta_1)$.

In summary, the polarizability derivatives of layer 1, layer 2 and layer 3 are $(\alpha'_1, \alpha'_2, \alpha'_3) = (\beta_1, -2\beta_1, \beta_1)$ for AA′A stacking; $(\beta_2, -2\beta_2, \beta_2)$ for ABA stacking; $(\beta_2, 0, -\beta_2)$ for ABC stacking; $(\beta_1, -\beta_1 + \beta_2, -\beta_2)$ for AA′B′ stacking; $(\beta_2, -\beta_2 - \beta_1, \beta_1)$ for A′B′B stacking. On the other hand, for weakly coupled layered materials such as GaSe, graphene, MoS$_2$, etc, there are two inter-layer shear modes (S$_1$ and S$_2$) for the trilayer, and their frequencies and eigenvectors (i.e., layer displacements) show little dependence on the stacking pattern. Therefore, regardless of the stacking detail in trilayer GaSe, the normalized displacements of layer 1, layer 2 and layer 3 are $(\Delta x_1, \Delta x_2, \Delta x_3) = \frac{1}{\sqrt{2}}(1, 0, -1)$ for the lower-frequency shear mode S$_1$, and $\frac{1}{\sqrt{1.5}}(0.5, -1, 0.5)$ for the higher-frequency shear mode S$_2$, according to the linear chain model. Based on the formula $\Delta \alpha = \sum_i \alpha'_i \cdot \Delta x_i$, we can subsequently obtain the polarizability changes by the shear vibrations as follows:

$\Delta\alpha(\text{AA′A, S}_1) = 0$; $\quad\quad\quad\quad\quad\quad \Delta\alpha(\text{AA′A, S}_2) = \sqrt{6}\beta_1$;

$\Delta\alpha(\text{ABA, S}_1) = 0$; $\quad\quad\quad\quad\quad\quad\ \Delta\alpha(\text{ABA, S}_2) = \sqrt{6}\beta_2$;

$\Delta\alpha(\text{ABC, S}_1) = \sqrt{2}\beta_2$; $\quad\quad\quad\quad\ \Delta\alpha(\text{ABC, S}_2) = 0$; $\quad\quad\quad\quad\quad\quad\quad$ (S14)

$\Delta\alpha(\text{AA′B′, S}_1) = \sqrt{0.5}(\beta_1 + \beta_2)$; $\quad \Delta\alpha(\text{AA′B′, S}_2) = \sqrt{1.5}(\beta_1 - \beta_2)$;

$$\Delta\alpha(A'B'B, S_1) = -\sqrt{0.5}(\beta_1 - \beta_2); \qquad \Delta\alpha(A'B'B, S_2) = \sqrt{1.5}(\beta_1 + \beta_2).$$

Since $I \propto \dfrac{n+1}{\omega}|\Delta\alpha|^2$, Raman intensities of the shear modes $S_1$ and $S_2$ at different stacking configurations in trilayer GaSe are the following:

$$I(AA'A, S_1) = 0; \qquad I(AA'A, S_2) \propto 6\dfrac{n_2+1}{\omega_2}|\beta_1|^2;$$

$$I(ABA, S_1) = 0; \qquad I(ABA, S_2) \propto 6\dfrac{n_2+1}{\omega_2}|\beta_2|^2;$$

$$I(ABC, S_1) \propto 2\dfrac{n_1+1}{\omega_1}|\beta_2|^2; \qquad I(ABC, S_2) = 0; \qquad (S15)$$

$$I(AA'B', S_1) \propto 0.5\dfrac{n_1+1}{\omega_1}|\beta_1+\beta_2|^2; \qquad I(AA'B', S_2) \propto 1.5\dfrac{n_2+1}{\omega_2}|\beta_1-\beta_2|^2;$$

$$I(A'B'B, S_1) \propto 0.5\dfrac{n_1+1}{\omega_1}|\beta_1-\beta_2|^2; \qquad I(A'B'B, S_2) \propto 1.5\dfrac{n_2+1}{\omega_2}|\beta_1+\beta_2|^2,$$

where $n_1$ and $\omega_1$ are the occupation number and frequency of the shear mode $S_1$, respectively; $n_2$ and $\omega_2$ are the occupation number and frequency of the shear mode $S_2$, respectively. According to the experimental data, $\omega_1 \approx 9.6$ cm$^{-1}$ and $\omega_2 \approx 16.6$ cm$^{-1}$, leading to $\dfrac{n_1+1}{\omega_1} = 2.30$ and $\dfrac{n_2+1}{\omega_2} = 0.78$ at room temperature. Therefore,

$$I(AA'A, S_1) = 0; \qquad I(AA'A, S_2) \propto 4.68|\beta_1|^2;$$

$$I(ABA, S_1) = 0; \qquad I(ABA, S_2) \propto 4.68|\beta_2|^2;$$

$$I(ABC, S_1) \propto 4.60|\beta_2|^2; \qquad I(ABC, S_2) = 0; \qquad (S16)$$

$$I(AA'B', S_1) \propto 1.15|\beta_1 + \beta_2|^2; \qquad I(AA'B', S_2) \propto 1.17|\beta_1 - \beta_2|^2;$$

$$I(A'B'B, S_1) \propto 1.15|\beta_1 - \beta_2|^2; \qquad I(A'B'B, S_2) \propto 1.17|\beta_1 + \beta_2|^2.$$

**Table S2.** Simulated intensity ratios of the spots in the HR-S/TEM images for AA′B′ and A′B′B stacking.

| Stacking sequences | Number of Ga atoms | Number of Se atoms | Intensity ratio |
|---|---|---|---|
| AA′B′ | 0 | 2 | 1 |
| | 2 | 2 | 1.51 |
| | 4 | 2 | 1.98 |
| A′B′B | 2 | 0 | 1 |
| | 2 | 2 | 1.65 |
| | 2 | 4 | 2.18 |

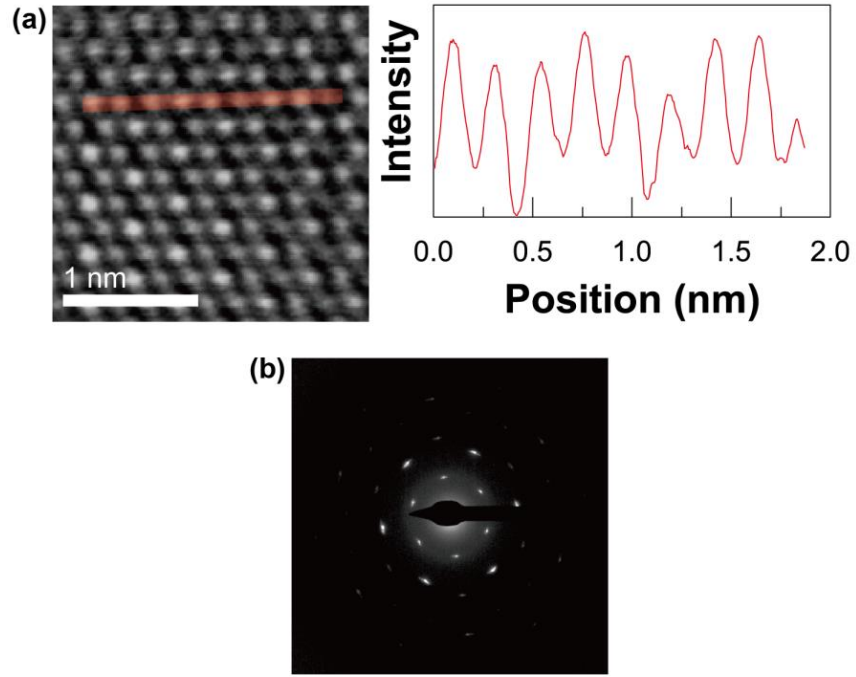

**Fig. S8.** (a) High-resolution TEM results from the surrounding region of the Type1 sample of Figure 5e showing an irregular intensity profile due to vacancies and (b) blurred diffraction pattern from the region showing poor crystallinity.